\documentclass[
superscriptaddress,
 amsmath,amssymb,
 aps,prb,twocolumn]{revtex4-2}

\usepackage{amsmath,amssymb}
\usepackage{amsfonts} 	
\usepackage{graphicx}   
\usepackage{verbatim}   
\usepackage{textgreek}
\usepackage{lipsum}
\usepackage{color}      
\usepackage{subfigure}  
\usepackage[normalem]{ulem} 
\usepackage[super]{nth} 
\usepackage{hyperref}   
\hypersetup{
    colorlinks=true,
    linkcolor=blue,
    urlcolor=black,
    citecolor=red
}
\usepackage{pbox}
\newcommand{\etaTETM}{\eta_{_{\text{TM/TE}}}}
\newcommand{\thetaTETM}{\theta_{_{\text{TM/TE}}}}

\begin{document}

\title{Chiral sensing with achiral anisotropic metasurfaces} 

\author{Sotiris Droulias}
\email{sdroulias@iesl.forth.gr}
\affiliation{Institute of Electronic Structure and Laser, FORTH, 71110, Heraklion, Crete, Greece}

\author{Lykourgos Bougas}
\email{lybougas@uni-mainz.de}
\affiliation{Institut f\"ur Physik, Johannes Gutenberg Universit\"at-Mainz, 55128 Mainz, Germany}


\begin{abstract}
Recently, we proposed a metasurface design for chiral sensing that (i) results in enhanced chiroptical signals by more than two orders of magnitude for ultrathin, subwavelength, chiral samples over a uniform and accessible area, (ii) allows for complete measurements of the total chirality (magnitude and sign of both its real and imaginary part), and (iii) offers the possibility for a crucial signal reversal (excitation with reversed polarization) that enables chirality measurements in an absolute manner, i.e., without the need for sample removal. Our design is based on the anisotropic response of the metasurface, rather than the superchirality of the generated near-fields, as in most contemporary nanophotonic-based chiral sensing approaches. Here, we derive analytically, and verify numerically, simple formulas that provide insight to the sensing mechanism and explain how anisotropic metasurfaces, in general, offer additional degrees of freedom with respect to their isotropic counterparts. We provide a detailed discussion of the key functionalities and benefits of our proposed design and we demonstrate practical measurement schemes for the unambiguous determination of an unknown chirality. Last, we provide the design principles towards broadband operation - from near-infrared to near-ultraviolet frequencies - opening the way for highly sensitive nanoscale chiroptical spectroscopy.

\end{abstract}
\keywords{chirality, absolute chiral sensing, metamaterials, optical rotation, circular dichroism, thin films}

\maketitle


\section{Introduction}
Chirality - a geometric property in which an object is non-superimposable with its mirror image - is an essential condition for the structural and functional diversity of biomolecules and many chemical compounds, which can exist in right- and left-handed forms known as enantiomers. Their functionality is often determined by their handedness and, hence, the ability to efficiently sense molecular chirality is of fundamental importance for many research disciplines and industries, such as the chemical, agricultural, and pharmaceutical\,\cite{barron2004,Busch2006}. \\
\indent A homogeneous natural optically active medium, e.g. a thin film consisting of chiral molecules, is described electromagnetically by the constitutive relations $\mathbf{D}=\epsilon_0 \epsilon\mathbf{E} –i(\kappa/c)\mathbf{H}$, $\mathbf{B}= \mu_0 \mu\mathbf{H}+i(\kappa/c)\mathbf{E}$\,\cite{Condon}, where $\epsilon$,$\mu$ are the relative permittivity and permeability ($\epsilon_0$,$\mu_0$ are the vacuum permittivity and permeability) and $c$ the vacuum speed of light; $\kappa$ is the chirality (also known as ‘Pasteur’\,\cite{Lindell1994}) parameter, which expresses the chiral molecular response and is, in general, complex: its magnitude, $|\kappa|$, is proportional to the sample's density (or its concentration in a solution), and its sign, sgn($\kappa$), depends on the handedness of the medium\,\cite{Condon}. Therefore, in order to be able to detect and distinguish enantiomers, a chiral sensing scheme should be sensitive to $|\kappa|$ and sgn($\kappa$), respectively. \\
\indent Among the most widely used techniques for chiral sensing are the polarimetric techniques of optical rotatory dispersion (ORD) and circular dichroism (CD)\,\cite{barron2004}, both routinely applied in industrial applications\,\cite{Busch2006}. Unfortunately, and despite their extensive use, the sensitivity limits of commercially available optical spectropolarimeters are at the $\sim0.1-1$\,mdeg levels\,\cite{Vaccaro2011}, which constrain their applicability for measurements of dilute samples (e.g., gas samples), and, particularly, of nanometer-scale thin films, where the chiroptical signals are typically $<100$\,\textmu deg (e.g., protein monolayers). \\
\indent On the other hand, nanophotonic approaches for chiral sensing have allowed for the measurement of chiroptical signals unattainable using traditional polarimetric techniques. Their principle of operation is based on the excitation of superchiral near fields, that is, fields with optical chirality density\,\cite{Tang2010} higher than that of circularly polarized light, and this approach is met in various schemes, such as propagating surface plasmons\,\cite{Pellegrini2017,Droulias2018ACSCHISPR}, plasmonic particles\,\cite{Govorov2010,Abdulrahman2012,Davis2013,Maoz2013}, chiral metamaterials\,\cite{Schaferling2012PRX,Tullius2015,Tullius2017,Zhang2017} and, recently, achiral metamaterials\,\cite{Mohammadi2018,Mohammadi2019,Guirado2020}. However, almost all recent nanophotonic approaches for chiral sensing have focused their efforts on performing and enhancing CD measurements, i.e., measurements of only the imaginary part of the chirality parameter, Im($\kappa$). On the contrary, the ability to detect Re($\kappa$) can be particularly useful in cases where Im($\kappa$) is weak, e.g. at frequencies far from the chiral molecular resonances. In addition, in most demonstrations and proposals, the employed nanophotonic systems have their own intrinsic chiroptical responses that contribute in the total measured CD signal (e.g., Refs.\,\cite{Tullius2015,Tullius2017,Zhao2017}), thus requiring separate measurements with and without the chiral inclusions to identify the chiroptical signal from the molecular system\,\cite{Zhao2017}. \\
\indent Recently, we proposed an achiral metasurface\,\cite{Droulias2020NL} that overcomes these limitations and (i) results in enhanced chiroptical signals by more than two orders of magnitude for ultrathin, subwavelength, chiral samples over a uniform and accessible area, (ii) allows for complete measurements of the total chirality (magnitude and sign of both its real and imaginary part), and (iii) offers the possibility for a crucial signal reversal (excitation with reversed polarization) that enables chirality measurements in an absolute manner, i.e., without the need for sample removal. Our scheme is based on the anisotropy of the metasurface, rather than the superchirality of the near-fields and, particularly in the context of possible future experiments and applications, it is necessary to introduce a general theoretical framework for anisotropic metasurfaces that provides further insight to the underlying sensing mechanism.\\
\indent In this work we provide a general theoretical model for the description of achiral anisotropic metasurfaces with chiral inclusions (i.e., a chiral molecule) that elucidates how aspects of chiral sensing are associated with the properties of the metasurfaces. For electrically thin metasurfaces, as considered in this work, the model is based on replacing the actual metasurface with a polarizable sheet that supports electric and magnetic surface currents, which are coupled via the chiral inclusions\,\cite{Droulias2020NL,Droulias2020PRB}. We show analytically, and verify numerically, that far-field measurements are proportional to the chirality parameter $\kappa$, thus justifying why the identification of both sgn($\kappa$) and $|\kappa|$ is possible\,\cite{Droulias2020NL}. Additionally, we show that these classes of metasurfaces have the ability to differentiate the real and imaginary part of $\kappa$, Re($\kappa$) and Im($\kappa$), respectively, contrary to most current approaches that are focused on enhancing CD signals and, as such, are expected to be sensitive to Im($\kappa$) \,\cite{Pellegrini2017, Nesterov2016, Zhao2017, Schaferling2012PRX, Mohammadi2018, Mohammadi2019, Graf2019, Solomon2019, Yao2019, Hu2020, Solomon2020}. We derive simple analytical formulas that provide insight to the mechanism associated with different excitation conditions (linearly/elliptically polarized waves) and explain how anisotropic metasurfaces can circumvent the trade-off met in isotropic metasurfaces, in which the far-field chiroptical response changes inversely with the transmitted power. Last, we demonstrate practical measurement schemes for the unambiguous determination of an unknown chirality, a crucial aspect in the design of future experiments, and provide the design principles towards chiroptical spectroscopy over different (broadband) spectral ranges.
\section{Theoretical formulation}
\noindent Based on the fact that chiral matter-wave interactions require a nonvanishing pseudoscalar product between $\mathbf{E}$ and $\mathbf{B}$, chiral sensing with nanophotonic systems is largely based on appropriately tailoring an electric and a magnetic mode of the system to be spatially and spectrally overlapped, and appropriately phased\,\cite{Schaferling2012OSA,Mohammadi2019}. The combined action of the two modes mediates the strong near-field coupling between the nanophotonic system and the chiral inclusion, which enables enhanced far-field signals from otherwise weak chirality values ($\kappa$). For the case of interest of electrically thin, sub-wavelength, systems such as metasurfaces, to obtain better insight into the underlying enhancement mechanism, we can eliminate the spatial dimension of the system and work with entirely time-dependent quantities. In essence, we can replace the metasurface by a thin polarizable sheet (zero thickness) and work with equivalent surface quantities (currents, conductivities).

\subsection{The conductivity tensor $\hat{\sigma}$}
\begin{figure}[t!]
\centering
		\includegraphics[width=\linewidth]{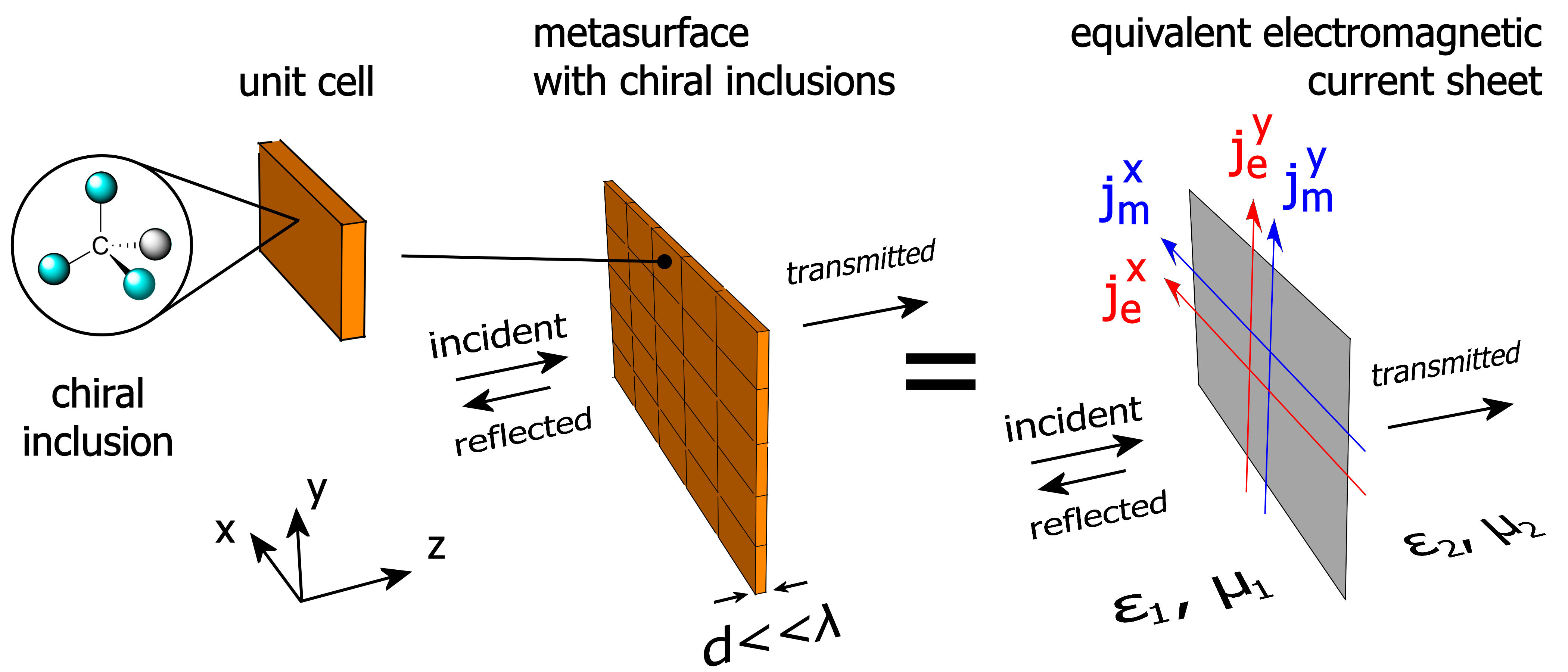}
	\caption{Schematic of an achiral anisotropic metasurface of thickness $d\ll\lambda$ ($\lambda$: wavelength) for enhanced chiral sensing of chiral inclusions. The metasurface supports modes with electric and magnetic moments with components along the $x,y$ directions that interact with the chiral inclusions. These moments are equivalently described by an infinite electromagnetic current sheet supporting electric currents $\mathbf{j}_e^x$, $\mathbf{j}_e^y$ and magnetic currents $\mathbf{j}_m^x$, $\mathbf{j}_m^y$, which are coupled via a magneto-electric conductivity.}	
    	\label{fig:fig01}
\end{figure}

\vspace{-3.5mm}
\noindent Let us replace the metasurface by a thin polarizable sheet that is located at $z = 0$ and extends on the $xy$-plane, as shown in Fig.\,\ref{fig:fig01}. The half-space for $z<0$ ($z>0$) is characterized by relative permittivity $\epsilon_1$ ($\epsilon_2$) and relative permeability $\mu_1$ ($\mu_2$). The sheet supports collinear electric and magnetic current densities, $\mathbf{j}_e$ and $\mathbf{j}_m$, respectively, which correspond to the electric and magnetic mode of the actual metasurface. A wave arriving at the sheet is partly reflected and partly transmitted, after exciting the surface current densities. Without loss of generality, we assume the anisotropic axes of the sheet to be the $x$- and $y$- axes, and, as such, $\mathbf{j}_e$ and $\mathbf{j}_m$ are, in general, different along these directions, and are related to the local (surface) fields $\mathbf{E}_{loc}$ and $\mathbf{H}_{loc}$ (at the location of the polarizable sheet, $z = 0$), as:
\begin{equation}
\label{eqn1}
\left(\begin{array}{c}
j^x_e \\
j^y_e \\
\hline
j^x_m \\
j^y_m
\end{array}\right)
=\overbrace{\left(\begin{array}{cc|cc}
\sigma^{xx}_{ee} & \sigma^{xy}_{ee} & \sigma^{xx}_{em} & \sigma^{xy}_{em} \\
\sigma^{yx}_{ee} & \sigma^{yy}_{ee} & \sigma^{yx}_{em} & \sigma^{yy}_{em} \\
\hline
\sigma^{xx}_{me} & \sigma^{xy}_{me} & \sigma^{xx}_{mm} & \sigma^{xy}_{mm} \\
\sigma^{yx}_{me} & \sigma^{yy}_{me} & \sigma^{yx}_{mm} & \sigma^{yy}_{mm}
\end{array}\right)}^{\hat{\sigma}}
\left(\begin{array}{c}
E^x_{loc} \\
E^y_{loc} \\
\hline
H^x_{loc} \\
H^y_{loc}
\end{array}\right),
\end{equation}
where $\hat{\sigma}$ is the conductivity tensor; its elements $\sigma^{\alpha \beta}_{i j}$ are surface (complex) conductivities, with the subscript $\{i,j\} = \{e,m\}$ denoting the electric `$e$' or magnetic `$m$' character of the associated conductivities and fields and the superscript $\{\alpha,\beta\} = \{x,y\}$ accounting for their associated components along the $x$-, $y$- directions. The electric and magnetic conductivities, $\sigma^{\alpha \beta}_{e e}$ and $\sigma^{\alpha \beta}_{m m}$, are measured in S and $\Omega$, respectively and the magneto-electric conductivities $\sigma^{\alpha \beta}_{e m}$, $\sigma^{\alpha \beta}_{m e}$ are dimensionless. In the most general case all elements of the conductivity tensor are nonzero. For achiral metasurfaces, as we consider here, and in the absence of any chiral inclusions, the two off-diagonal $2 \times 2$ blocks are zero (all $\sigma^{\alpha \beta}_{i j}$ with $i\neq j$). Additionally, each of the two diagonal $2 \times 2$ blocks can be expressed in diagonal form, leaving only four quantities necessary to describe the metasurface, that is $\sigma^{x x}_{e e}$, $\sigma^{y y}_{e e}$, $\sigma^{x x}_{m m}$ and $\sigma^{y y}_{m m}$.
The presence of chirality ($\kappa \neq 0$) introduces magneto-electric coupling, thus rendering nonzero the components $\sigma^{\alpha \beta}_{i j}$ with $i\neq j$. We may write (i) $\sigma^{\alpha \beta}_{e m} = -\sigma^{\alpha \beta}_{m e} \equiv \sigma^{\alpha \beta}_c$ for $\alpha=\beta$ and (ii) $\sigma^{\alpha \beta}_{e m} = \sigma^{\alpha \beta}_{m e} = 0$ for $\alpha\neq\beta$, generalizing the respective dependencies examined previously for isotropic metasurfaces in Ref.\,\cite{Droulias2020PRB}, and, hence, the conductivity tensor takes the form: 
\begin{equation}
\label{eqn2}
\hat{\sigma}=\left(\begin{array}{cccc}
\sigma^{xx}_{ee} & 0 & \sigma^{xx}_c & 0 \\
0 & \sigma^{yy}_{ee} & 0 & \sigma^{yy}_c \\
-\sigma^{xx}_c & 0 & \sigma^{xx}_{mm} & 0\\
0 & -\sigma^{yy}_c & 0 & \sigma^{yy}_{mm}
\end{array}\right).
\end{equation}
This is the general form of $\hat{\sigma}$, which we will examine throughout this work. Next, without loss of generality, we will examine metasurfaces with chiral inclusions embedded in a uniform environment, that is, with $\epsilon_1=\epsilon_2\equiv\epsilon$ and $\mu_1=\mu_2\equiv\mu$. For the general case where $\epsilon_1 \neq \epsilon_2$, $\mu_1 \neq \mu_2$, see Supplemental Material (SM) for the full analytical expressions.\\
\indent To understand the magneto-electric coupling, as expressed via $\sigma^{xx}_c$, $\sigma^{yy}_c$, we can generalize the result derived previously in Ref.\,\cite{Droulias2020PRB} for isotropic achiral metasurfaces with chiral inclusions. There, we show that the magnetoelectric conductivity is proportional to three quantities, namely (i) the chirality parameter $\kappa$ of the chiral inclusions, (ii) an overlap integral of the fields of the electric and magnetic mode, and (iii) the product of their individual conductivities. In the case of anisotropic metasurfaces, a similar derivation leads us to write an equivalent expression along each individual $x$-, $y$- direction, as: 
\begin{equation}
\label{eqn3}
\sigma^{\alpha \alpha}_c = \kappa \cdot C_0 \cdot s^{\alpha \alpha}_{ee}s^{\alpha \alpha}_{mm},  \quad a = \{x,y\}
\end{equation}
where $C_0$ is a constant resulting from the overlap integral of the fields. 

\subsection{Scattering amplitudes and retrieval of conductivities}\vspace{-3.5mm}
\noindent To examine the effects of chirality on a wave impinging on a system that is described electromagnetically by the conductivity tensor of Eq.\@(\ref{eqn2}), we solve Maxwell's equations for the thin current sheet with the appropriate boundary conditions, which are formulated as $\mathbf{n} \times (\mathbf{E}_2 – \mathbf{E}_1) = –\mathbf{j}_m$, $\mathbf{n} \times (\mathbf{H}_2 – \mathbf{H}_1) = +\mathbf{j}_e$, with $\mathbf{n}$ being the surface normal of the current sheet pointing from region 1 to region 2 (Fig.\,\ref{fig:fig01}). It is also convenient to introduce the dimensionless conductivities $s^{\alpha \alpha}_{e e} = \zeta \sigma^{\alpha \alpha}_{e e}/2$ and $s^{\alpha \alpha}_{m m} = \sigma^{\alpha \alpha}_{m m}/2\zeta$, where $\alpha = \{x,y\}$ ($\sigma^{\alpha \alpha}_c$ is dimensionless by definition), and $ \zeta = \sqrt{\mu_0 \mu/\epsilon_0 \epsilon}$ is the wave impedance of the surroundings. Solving for $x$\,- and $y$\,-linearly polarized incident waves, we find that the transmission and reflection amplitudes are expressed in terms of the surface conductivities as:   
\begin{subequations}
\label{eq:eqn4}
\begin{gather}
        t_{xx}=\frac{1-s^{xx}_{ee}s^{yy}_{mm}}{(1+s^{xx}_{ee})(1+s^{yy}_{mm})},
        \label{eq:set4A} \\
        t_{yy}=\frac{1-s^{yy}_{ee}s^{xx}_{mm}}{(1+s^{yy}_{ee})(1+s^{xx}_{mm})},	 
        \label{eq:set4B} \\
        t_c = \frac{\sigma^{xx}_c}{2(1+s^{xx}_{ee})(1+s^{xx}_{mm})} +\frac{\sigma^{yy}_c}{2(1+s^{yy}_{ee})(1+s^{yy}_{mm})},
        \label{eq:set4C} \\
        r_{xx}=\frac{s^{yy}_{mm}-s^{xx}_{ee}}{(1+s^{xx}_{ee})(1+s^{yy}_{mm})}, 
        \label{eq:set4D} \\
        r_{yy}=\frac{s^{xx}_{mm}-s^{yy}_{ee}}{(1+s^{yy}_{ee})(1+s^{xx}_{mm})},
        \label{eq:set4E} \\
        r_c = \frac{\sigma^{xx}_c}{2(1+s^{xx}_{ee})(1+s^{xx}_{mm})} -\frac{\sigma^{yy}_c}{2(1+s^{yy}_{ee})(1+s^{yy}_{mm})},
        \label{eq:set4F}
\end{gather}
\end{subequations}
or, solving in terms of the conductivities:
\begin{subequations}
\label{eq:eqn5}
\begin{gather}
        s^{xx}_{ee}=\frac{1-r_{xx}-t_{xx}}{1+r_{xx}+t_{xx}},
        \label{eq:set5A} \\
        s^{yy}_{ee}=\frac{1-r_{yy}-t_{yy}}{1+r_{yy}+t_{yy}}, \label{eq:set5B} \\
        s^{xx}_{mm}=\frac{1+r_{yy}-t_{yy}}{1-r_{yy}+t_{yy}}, \label{eq:set5C} \\
        s^{yy}_{mm}=\frac{1+r_{xx}-t_{xx}}{1-r_{xx}+t_{xx}}, \label{eq:set5D} \\
        \sigma^{xx}_c=\frac{4(t_c+r_c)}{(1+r_{xx}+t_{xx})(1-r_{yy}+t_{yy})},	 \label{eq:set5E} \\
        \sigma^{yy}_c=\frac{4(t_c-r_c)}{(1-r_{xx}+t_{xx})(1+r_{yy}+t_{yy})},	 \label{eq:set5F}
\end{gather}
\end{subequations}
where the subscripts in $t_{out,inc}$, $r_{out,inc}$ denote the output and incident $E$-field polarization, respectively, and we define $r_{xx},t_{xx},r_{yy},t_{yy}$ and $t_c\equiv t_{xy} = -t_{yx}, r_c\equiv r_{xy} = r_{yx}$ as the co- and cross- polarized scattering amplitudes, respectively. 
\begin{figure*}[t!]
\centering
		\includegraphics[width=\linewidth]{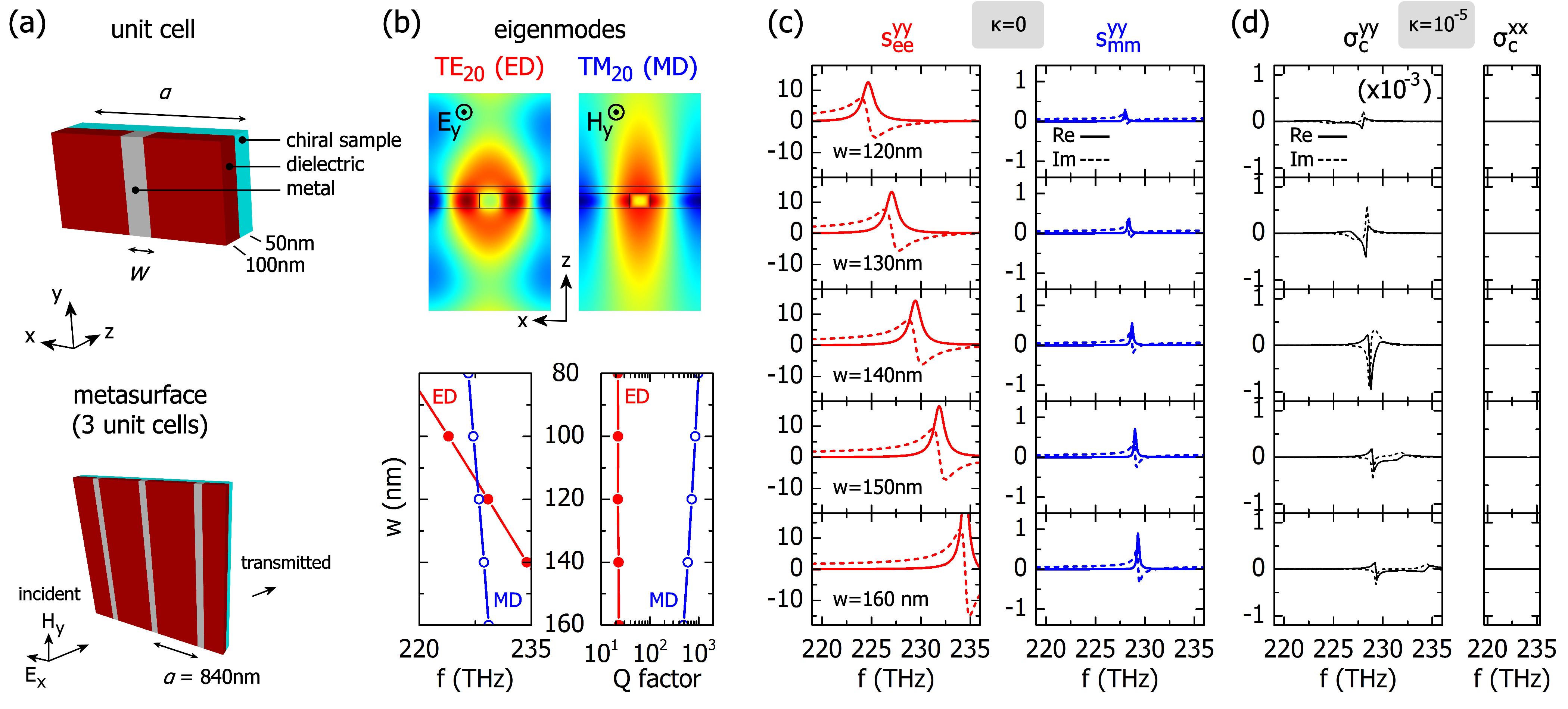}
	\caption{Anisotropic achiral metasurface for enhanced chiral sensing. (a) Schematic of a single unit cell (top) and of the metasurface (bottom). (b) $\textit{Top panel}$: field distribution of the electric-type (TE$_{20}$ or ED) and magnetic-type (TM$_{20}$ or MD) modes. $\textit{Bottom panel}$: spectral tuning (left panel) and $Q$ factors (right panel) of TE$_{20}$ and TM$_{20}$ as a function of the metal width, $w$. (c) Retrieved conductivities $\sigma^{y y}_{e e}$, $\sigma^{y y}_{m m}$ as a function of the metal width, $w$, for $\kappa = 0$. (d) Retrieved conductivies $\sigma^{y y}_c$, $\sigma^{x x}_c$ as a function of the metal width, $w$, for $\kappa = 10^{-5}$.}	
    	\label{fig:fig02}
\end{figure*}
To derive Eqs.\@(\ref{eq:set4A})-(\ref{eq:set4F}) and Eqs.\@(\ref{eq:set5A})-(\ref{eq:set5F}) we make the approximation $\sigma^{\alpha \alpha}_c \ll s^{\alpha \alpha}_{e e}, s^{\alpha \alpha}_{m m}$, as in \cite{Droulias2020PRB}, thus eliminating any $(\sigma^{\alpha \alpha}_c)^2$ term; this approximation is valid, as long as the magneto-electric coupling is perturbative. Importantly, because $\kappa$ is dispersive, $\sigma^{\alpha \alpha}_c$ exhibits the dispersive features of both the chiral inclusions and the metasurface\,\cite{Droulias2020PRB}. Therefore, in order to isolate the contribution from the metasurface we will start with a constant $\kappa$, which is a reasonable approximation when studying metasurfaces with resonances of much narrower linewidth than typically that of the chiral inclusions. With Eqs.\@(\ref{eq:set5A})-(\ref{eq:set5F}), we can now use $r_{\alpha \beta}$, $t_{\alpha \beta}$ from simulations or experiments with the metasurface to retrieve the surface conductivities $s^{\alpha \alpha}_{e e}$, $s^{\alpha \alpha}_{m m}$, $\sigma^{\alpha \alpha}_c$ of its equivalent sheet model. \\
\indent As an example, let us consider an anisotropic metasurface that supports resonant conductivities only along the $y$-axis, similar to the one examined in Ref.\,\cite{Droulias2020NL}. We focus on design parameters for chiral sensing in the visible and near-infrared, and in particular, we choose 1300\,nm ($\sim230$\,THz). The metasurface is composed of a 100\,nm thin dielectric slab on which a 50\,nm chiral layer is placed [see Fig.\,\ref{fig:fig02}(a)]. The chiral layer has refractive index $n = 1.5-0.001i$ and chirality parameter $\kappa = 10^{-5}$ (realistic chirality parameter value of, e.g., aqueous solutions of monosaccharides\,\cite{Sofikitis2014,Bougas2015} or biomolecules\,\cite{Abdulrahman2012,Kelly2018,Guirado2020}). 
The dielectric slab has refractive index $n = 3.4$ (e.g., Si) and is periodically interrupted by metallic wires of the same thickness (100\,nm) and width $w$, with periodicity $\alpha = 840$\,nm; the metal is a Drude silver (Ag) of permittivity based on Johnson and Christy data\,\cite{JohnsonChristy}. The whole system is placed on a glass substrate of refractive index $n = 1.5$ and the space on the opposite side (adjacent to the chiral layer) is index-matched with the substrate. The dielectric slab supports TE (components $H_x, E_y, H_z$) and TM (components $E_x, H_y, E_z$) waveguide modes, which we use to implement the electric/magnetic-moment pair. This is achieved with the periodically distributed metallic wires, the role of which is to spatially quantize the TE/TM waveguide modes and provide discrete sets of resonant states. The presence of the metallic wires disturbs the symmetry of the fields across each unit cell, essentially resulting in residual moments in the dominant field components of each mode, i.e., electric (magnetic) dipole moment in the $E_y$ ($H_y$) component of the TE (TM) mode. The width, $w$, of the metallic wires is used for tuning the spectral separation between the two modes, which we label as TE$_{20}$ (electric dipole or ED) and TM$_{20}$ (magnetic dipole or MD), in accordance with our previous work\,\cite{Droulias2020NL}. The spatial distribution of their dominant field components $E_y$ and $H_y$, respectively, is shown in Fig.\,\ref{fig:fig02}(b), where also their frequency tuning and individual $Q$ factors are shown. For our numerical simulations we perform full-wave vectorial Finite Element Method (FEM) simulations, with the commercial software COMSOL Multiphysics. \\
\indent To examine the response of the metasurface, first, with $\kappa = 0$, we send separately $x$\,- and $y$\,-linearly polarized waves, to excite the system along its anisotropic axes. A $x$- ($y$-) polarized incident wave excites the TM$_{20}$ (TE$_{20}$) mode, which cannot couple to the orthogonal TE$_{20}$ (TM$_{20}$) mode, resulting in only the co-polarized scattering amplitudes $t_{xx}, t_{yy}, r_{xx}, r_{yy}$ being nonzero. We then use the latter with Eqs.\@(\ref{eq:set5A})-(\ref{eq:set5D}) to retrieve $s^{x x}_{e e}$, $s^{y y}_{m m}$, $s^{y y}_{e e}$, $s^{x x}_{m m}$. Next, we set $\kappa = 10^{-5}$ and repeat the two experiments: illumination with $x$- ($y$-) polarized wave again leads to the excitation of the TM$_{20}$ (TE$_{20}$) mode, however, due to chirality, the two modes can couple, giving rise to the cross-polarized terms $t_c$, $r_c$ and magneto-electric conductivities $\sigma^{xx}_c$, $\sigma^{yy}_c$.
Note that, for the chosen (weak) $\kappa = 10^{-5}$, the co-polarized scattering amplitudes $(t_{xx}, t_{yy}, r_{xx}, r_{yy})$ remain practically unchanged, and so do the conductivities $s^{x x}_{e e}$, $s^{y y}_{m m}$, $s^{y y}_{e e}$, $s^{x x}_{m m}$ (retrieved in the previous step). In Fig.\,\ref{fig:fig02}\,(c) we show $s^{yy}_{ee}$, $s^{yy}_{mm}$ for metal widths, $w$, ranging from 120\,nm to 160\,nm (see SM for comparison with $s^{xx}_{ee}$, $s^{xx}_{mm}$), which we use with Eqs.\@(\ref{eq:set5E})-(\ref{eq:set5F}) to retrieve $\sigma^{xx}_c$, $\sigma^{yy}_c$, as shown in Fig.\,\ref{fig:fig02}\,(d). The form of $\sigma^{xx}_c$, $\sigma^{yy}_c$ also verifies the strong inherent anisotropy of the metasurface: as we tune the metal width $w$, we detune the resonant frequencies of $s^{yy}_{ee}$, $s^{yy}_{mm}$, leading to a distinct change in $\sigma^{yy}_c$, while $\sigma^{xx}_c$ remains practically zero.

\subsection{Retrieval of $\kappa$ in terms of the conductivities}
\vspace{-3.5mm}
\begin{figure}[t!]
\centering
		\includegraphics[width=0.98\linewidth]{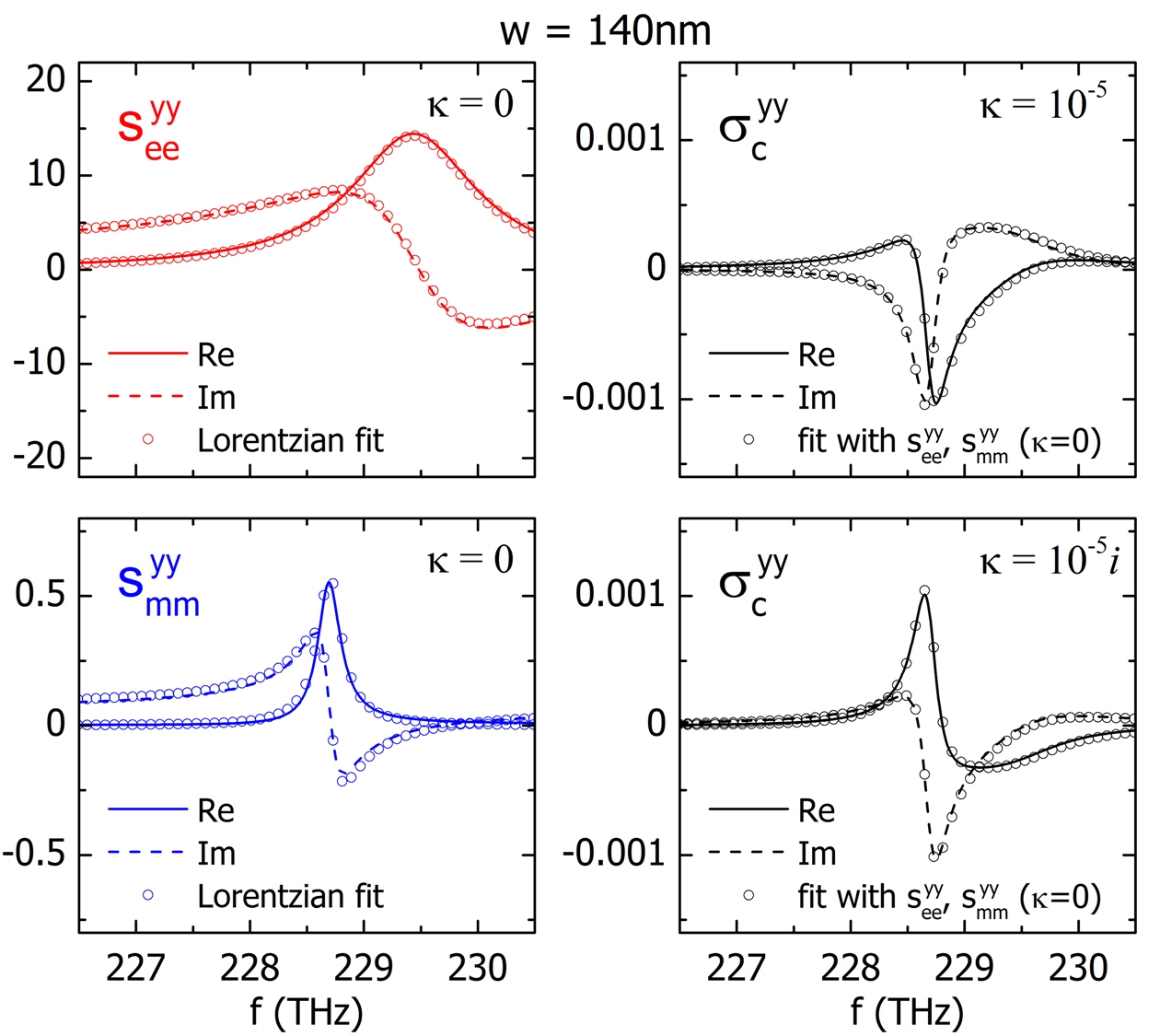}
	\caption{Numerically retrieved conductivities for the achiral anisotropic metasurface of Fig.\,\ref{fig:fig02} with $w = 140$\,nm. $\textit{Left column}$: electric $s^{y y}_{e e}$ and magnetic $s^{y y}_{m m}$ conductivity for $\kappa = 0$ and fit with Lorentzian function (open circles). $\textit{Right column}$: magneto-electric conductivity $\sigma^{y y}_c$ for $\kappa = 10^{-5}$ (top panel) and $\kappa = 10^{-5}i$ (bottom panel) and analytical fit with Eq.\,\ref{eq:eqn5} (open circles), using $s^{y y}_{e e}$, $s^{y y}_{m m}$ from the left column.}	
    	\label{fig:fig03}
\end{figure}
\noindent We can now use the retrieved conductivities to find the chirality parameter $\kappa$ of an unknown inclusion. In Eq.\@(\ref{eqn3}) we showed that $\sigma_c$ is proportional to $\kappa$, and, therefore, we need first to verify that the numerically retrieved conductivities satisfy this simple linear relation.\\
\indent We start by fitting a Lorentzian function to $s^{yy}_{ee}$, $s^{yy}_{mm}$:
\begin{equation}
\label{eqn6}
s^{yy}_{ee/mm}(\omega)=\frac{ia_{e/m}\omega}{\omega^2_{e/m}-\omega^2+i\gamma_{e/m}\omega}+i\beta_{e/m}\omega,
\end{equation}
where the subscript $e/m$ denotes the respective parameters for the electric/magnetic conductivity (the modification of the Lorentzian is because the current is the time derivative of the dielectric polarization). In Fig.\,\ref{fig:fig03}, left column, we plot $s^{yy}_{ee}$ and $s^{yy}_{mm}$ for $w = 140$\,nm [repeated from Fig.\,\ref{fig:fig03}(c), $\kappa = 0$] and their Lorentzian fits (open circles) with parameters $\omega_e = 2\pi \times 229.44$ THz, $\omega_m = 2\pi \times 228.7$\,THz, $\gamma_e = 2\pi \times 1.3$\,THz, $\gamma_m = 2\pi \times 0.26$ THz, $\alpha_e = 2\pi \times 18.5$\,THz, $\alpha_m = 2\pi \times 0.15$\,THz, $\beta_e = (2\pi)^{-1} \times 5.8$\,fs and $\beta_m = (2\pi)^{-1} \times 0.3$\,fs. In the right column of Fig.\,\ref{fig:fig03}, top panel, we plot the numerically retrieved $\sigma^{yy}_c$ for $\kappa = 10^{-5}$ (black lines) and its analytical fit (open circles) with Eq.\@(\ref{eqn3}), i.e., $\sigma^{fit}_c(\omega) = \kappa C_0 (s^{yy}_{ee}-i\beta _e \omega)(s^{yy}_{mm}-i\beta _m \omega)$. In this expression $\kappa = 10^{-5}$ (i.e., the chirality parameter used in the simulations), $C_0 = –23$ a constant which we use for the fitting, and $\beta_e = (2\pi)^{-1} \times 5.8$\,fs and $\beta_m = (2\pi)^{-1} \times 0.3$\,fs are the parameters used previously in the Lorentzians; $s^{yy}_{ee}$, $s^{yy}_{mm}$ are the retrieved conductivities. The reason for subtracting the terms related to $\beta_e, \beta_m$ is that, besides the electric and magnetic mode, in the actual metasurface there are other modes at nearby frequencies contributing a background, while in Eq.\@(\ref{eqn3}) $s^{yy}_{ee}, s^{yy}_{mm}$ are the resonant responses of exactly one electric and one magnetic mode, respectively. The agreement between the numerically retrieved conductivity $\sigma^{yy}_c$ and its analytical fit confirms the simple functional form of the magneto-electric coupling, as described by Eq.\@(\ref{eqn3}) and justifies the approximation $\sigma^{\alpha \alpha}_c\ll s^{\alpha \alpha}_{ee}$, $s^{\alpha \alpha}_{mm}$, which is also evident in the relative values of $\sigma_c$ and $s_{ee}$, $s_{mm}$, shown in Fig.\,\ref{fig:fig03}. Next, we repeat the simulations with $\kappa = 10^{-5}i$, a purely imaginary chirality parameter, and we plot the numerically retrieved $\sigma^{yy}_c$ in the right column of Fig.\,\ref{fig:fig03}, bottom panel (black lines). Using $C_0$, $\beta_e$, $\beta_m$ from the previous step we calculate the new analytical fit of $\sigma^{yy}_c$, which we overlap in the same plot as open circles. We can also go beyond the specific configuration and change the metal width $w$; in this case we retrieve the new $s^{yy}_{ee}$, $s^{yy}_{mm}$, which we use to evaluate $\sigma^{fit}_c$ as a function of $w$. For metal widths ranging from 120\,nm to 160\,nm and various choices of $\kappa$ we find excellent agreement between the numerically retrieved $\sigma^{yy}_c$ and its fit with $\sigma^{fit}_c$, as predicted by Eq.\@(\ref{eqn3}) (see SM). We emphasize that in all mentioned changes we use the above (fixed) values of $C_0$, $\beta_e$, $\beta_m$. \\
\indent The above conclusions verify the linearity of $\sigma^{yy}_c$ in $\kappa$, essentially enabling us to determine the chirality parameter $\kappa_A$ of an unknown chiral inclusion (A), using another inclusion of known chirality, $\kappa_B$, as reference. To see how this is possible, we need to take into account the fact that, as we showed in the previous section, $s^{yy}_{ee}$, $s^{yy}_{mm}$ remain practically unchanged as we change the chirality parameter $\kappa$ of the chiral inclusion. Therefore, after we separately apply Eq.\@(\ref{eqn3}) for both known and unknown inclusions, A and B, respectively, we can take into account that $\sigma_c \propto \kappa$, to eliminate the common $s^{yy}_{ee}$, $s^{yy}_{mm}$ terms. Then, the unknown chirality will be simply given by $\kappa_A = \kappa_B(\sigma^{yy}_{c,A}/\sigma^{yy}_{c,B})$, where $\sigma^{yy}_{c,A}, \sigma^{yy}_{c,B}$ are the numerically retrieved conductivies of inclusions A and B, respectively.

\section{Complete measurement with TM/TE linearly polarized beams}
\noindent To deduce the chirality parameter $\kappa$ of an unknown chiral inclusion according to the procedure outlined in the previous section, we used all reflection and transmission amplitudes. However, in most relevant experiments \,\cite{Mohammadi2018, Mohammadi2019, Graf2019, Solomon2019, Yao2019, Hu2020, Solomon2020, Droulias2020NL, Guirado2020, Zhao2017} chiral sensing relies on measurements in transmission only, naturally raising the question of how this can be also implemented in the framework of anisotropic metasurfaces. As we show next, this is possible if we analyze the polarization of the transmitted wave in terms of its rotation $\theta$ and ellipticity $\eta$, using the transmission amplitudes given by Eqs.\@(\ref{eq:set4A})-(\ref{eq:set4C}).
\subsection{Chiroptical signals in transmission}
\vspace{-3.5mm}
\begin{figure*}[t!]
\centering
		\includegraphics[width=0.98\linewidth]{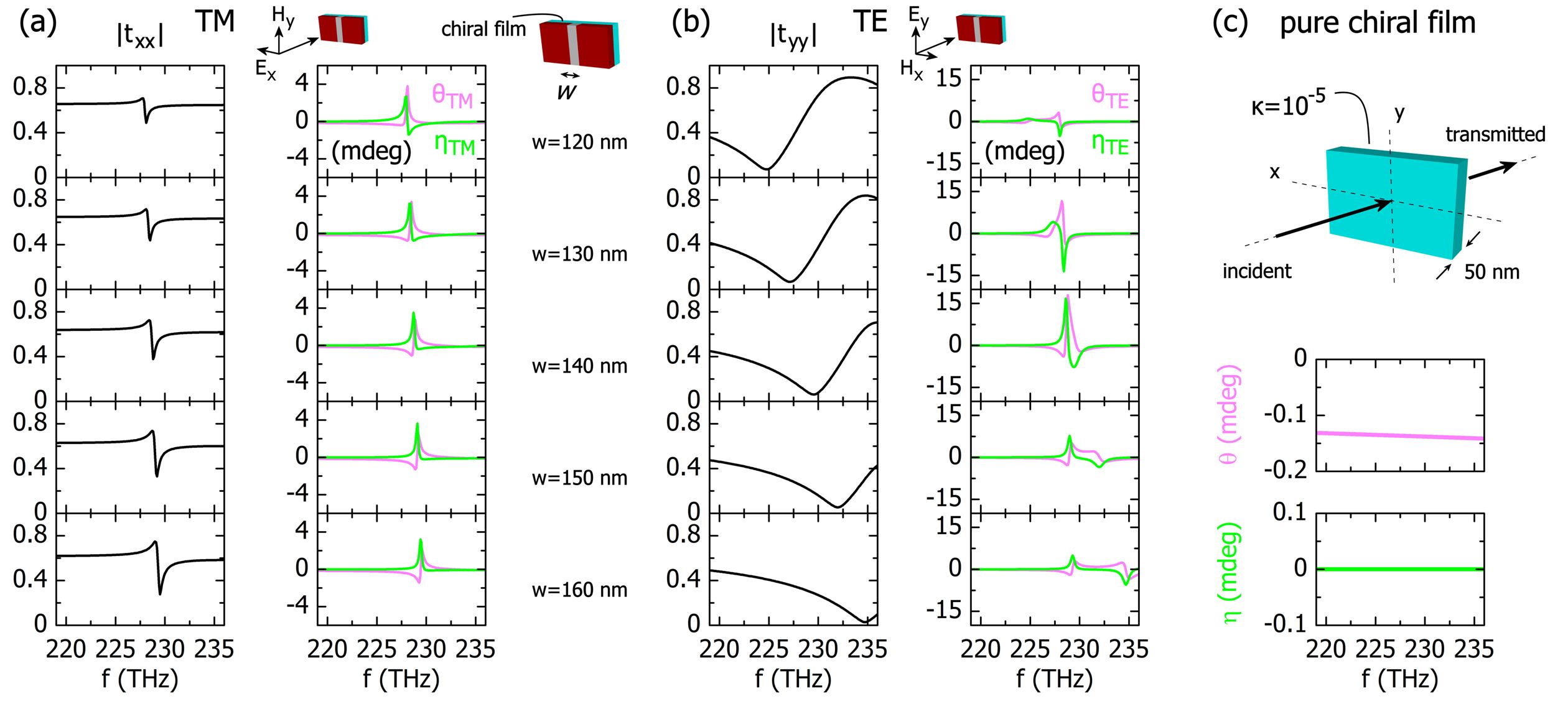}
	\caption{Response of anisotropic achiral metasurface with chiral inclusions, e.g., a chiral film, for enhanced chiral sensing. Absolute values of co-transmission amplitudes together with chiroptical signals $\theta$ (magenta lines) and $\eta$ (green lines) for $\kappa = 10^{-5}$ (purely real) under (a) TM-illumination ($x$-polarized incident $E$-field) and (b) TE-illumination ($y$-polarized incident $E$-field). In (c) the equivalent signals are shown for the pure chiral film, i.e. with the metasurface removed. The anisotropy offers an additional degree of freedom with respect to the available transmitted photons ($|t_{xx}|^2$ or $|t_{yy}|^2$).}	
    	\label{fig:fig04}
\end{figure*}
\noindent With the aid of the Stokes parameters (see SM) we find that the chiroptical signals $\theta$ and $\eta$ can be directly expressed in terms of $t_{xx}$, $t_{yy}$, and $t_c$ of Eqs.\@(\ref{eq:set4A})-(\ref{eq:set4C}) as:
\begin{subequations}
\label{eq:eqn7}
\begin{align}
        \thetaTETM={\rm{Re}}(-\frac{t_c}{t_{xx/yy}}),
        \label{eq:set7A} \\
        \etaTETM={\rm{Im}}(-\frac{t_c}{t_{xx/yy}}),
        \label{eq:set7B}
\end{align}
\end{subequations}
where we have used $t_c\ll t_{xx}, t_{yy}$ to approximate $\tan^{-1}(\Phi)\sim\Phi$. We refer to the illumination with $\mathbf{E}\parallel\hat{x}$ $(\mathbf{E}\parallel\hat{y})$ as TM (TE) illumination, to emphasize the fact that this particular polarization directly couples with the TM$_{20}$ (TE$_{20}$) mode, as also indicated in the subscripts of $\theta$, $\eta$.

From Eqs.\@(\ref{eq:set7A}),(\ref{eq:set7B}) we see that the chiroptical signals $\theta$, $\eta$ are in general different among the two polarizations. This is in contrast to the case for isotropic metasurfaces (Ref.\,\cite{Droulias2020PRB}) where the chiroptical signals are the same for both $x$-, $y$- linearly polarized waves and inversely proportional to the co- transmission amplitude $t\equiv t_{xx}=t_{yy}$. This leads to an important advantage: because anisotropy allows for the independent control of the conductivities along the different $x,y$ axes, we can independently tune the co- and cross-transmission amplitudes and, likewise, the chiroptical signals. For example, notice in Eq.\@(\ref{eq:set4A}) how $t_{xx}$ changes only with $s^{xx}_{ee}$, $s^{yy}_{mm}$ whereas all $s^{xx}_{ee}$, $s^{yy}_{mm}$, $s^{yy}_{ee}$, $s^{xx}_{mm}$ are involved in the cross-polarized term $t_c$; the additional conductivities $s^{yy}_{ee}$, $s^{xx}_{mm}$ can be used to tune $\theta$, $\eta$ independently from the transmittance $t_{xx}$, which remains unchanged. This gives an additional degree of freedom to maximize the chiroptical signals for a certain achievable transmittance. Similarly, the same conclusions hold for simpler systems, e.g., with resonant conductivities only along the $y$-axis as in our example; upon TM-illumination, $t_{xx}$ is tuned via $s^{yy}_{mm}$, while $t_c$ is tuned via both $s^{yy}_{mm}$ and $s^{yy}_{ee}$. \\
\indent To demonstrate the above findings we use a constant chirality parameter $\kappa = 10^{-5}$, as previously considered in Fig.\,\ref{fig:fig02}. In Figs.\,\ref{fig:fig04}\,(a),\,(b) we show the system’s response for both TM- and TE- illumination, respectively, as a function of the the metal width $w$. These results demonstrate the additional degree of freedom offered by the system's anisotropy in terms of the transmitted power ($|t_{xx}|^2$ or $|t_{yy}|^2$) and the attainable chiroptical signals. For example, for this particular system, one can choose between strong chiroptical signals $\theta_{_{\rm{TE}}}$, $\eta_{_{\rm{TE}}}$ on a relatively weak output beam [Fig.\,\ref{fig:fig04}\,(b)] or somewhat weaker chiroptical signals $\theta_{_{\rm{TM}}}$, $\eta_{_{\rm{TM}}}$ on a significantly stronger output beam [Fig.\,\ref{fig:fig04}\,(a)]. We emphasize that this cannot be achieved with isotropic systems\,\cite{Droulias2020PRB}. For both choices of illumination, the chiroptical signals are 30-100 times stronger than those from the chiral inclusions alone, as for a 50\,nm thin chiral film with $\kappa = 10^{-5}$ we find $\theta = 0.14$\,mdeg and $\eta = 0$ at 230\,THz, in the absence of the metasurface [Fig.\,\ref{fig:fig04}\,(c)]. The numerically calculated chiroptical signals are also in agreement with their analytical form given by Eqs.\@(\ref{eq:set7A}),(\ref{eq:set7B}), as we verify with separately calculating the ratios $t_c/t_{xx}$ and $t_c/t_{yy}$ and comparing their real and imaginary parts with $\thetaTETM$, $\etaTETM$ (see SM).\\
\indent Last, we note here that due to the linearity of $s^{xx}_c$, $s^{yy}_c$ in $\kappa$ [see Eq.\@(\ref{eqn3})], we find from Eq.\@(\ref{eq:set4C}) that $t_c\propto \kappa$. Therefore, Eqs.\@(\ref{eq:set7A}),(\ref{eq:set7B}) demonstrate that the far-field chiroptical signals $\thetaTETM$, $\etaTETM$ are directly proportional to $\kappa$ and justify why measurements of $\theta$, $\eta$ are expected to be sensitive to both $|\kappa|$ and sgn($\kappa$), and to both Re($\kappa$) and Im($\kappa$). This can be easily seen if we allow $\kappa$ to be dispersive, as we do next.

\subsection{Determination of unknown chirality using chiroptical signals in transmission}
\begin{figure}[t!]
\centering
\includegraphics[width=0.98\linewidth]{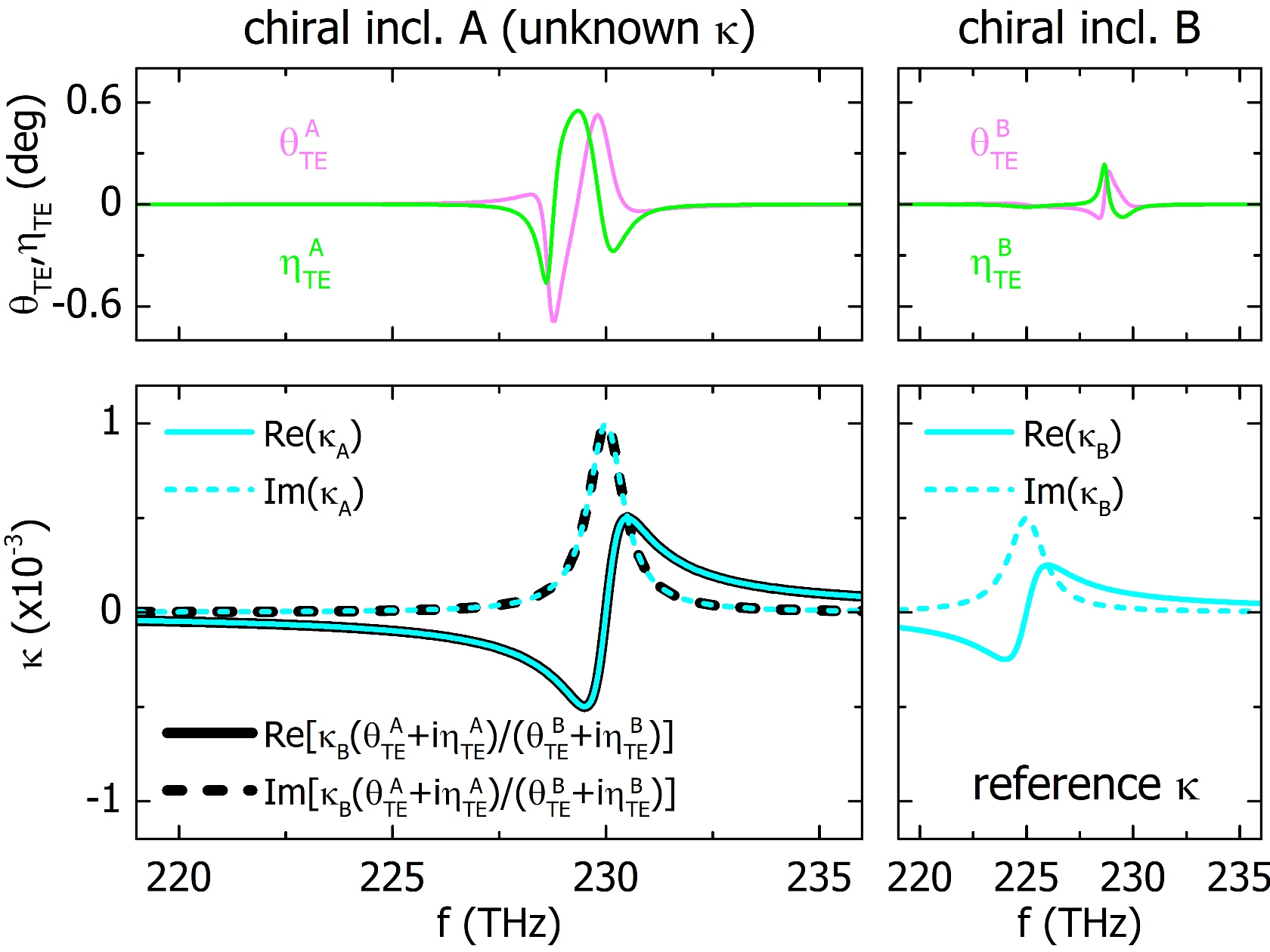}
	\caption{Retrieval of the chirality parameter $\kappa_A$ of an unknown inclusion A, using a reference chirality parameter $\kappa_B$ of inclusion B, and the chiroptical signals $\theta_{TE}$, $\eta_{TE}$ in transmission. $\textit{Top row}$: numerically calculated chiroptical signals $\theta_{TE}$, $\eta_{TE}$ for each inclusion, for the achiral anisotropic metasurface of Fig. 2 with $w = 140 nm$. $\textit{Bottom row}$: chirality parameters $\kappa_A$, $\kappa_B$ of the two inclusions (cyan lines) and retrieved $\kappa_A$ (black lines) of inclusion A.}	
    	\label{fig:fig05}
\end{figure}

\vspace{-3.5mm}
\noindent According to the results of Eqs.\@(\ref{eq:set7A}),(\ref{eq:set7B}), $\thetaTETM$ and $\etaTETM$ are the real and imaginary part, respectively, of the same quantity, i.e., $–t_c/t_{xx/yy}$. Therefore, we may combine the two expressions in each of Eqs.\@(\ref{eq:set7A}),(\ref{eq:set7B}) and solve for $t_c$, to obtain:
\begin{equation}
\label{eqn8}
t_c=-(\thetaTETM+i\etaTETM)\cdot t_{xx/yy}.
\end{equation}
The result of Eq.\@(\ref{eqn8}) is of great importance, as it enables us to unambiguously determine an unknown chirality solely from the chiroptical signals $\theta$, $\eta$ in transmission. To see how this is possible, let us consider two chiral inclusions, A and B, which we probe with our system under TM/TE-illumination. The chiroptical signals $\thetaTETM^A$, $\etaTETM^A$, and $\thetaTETM^B$, $\etaTETM^B$, obtained either from simulations or experiments using inclusions A and B, respectively, correspond to the individual transmission amplitudes $t^A_c$ and $t^B_c$, according to Eq.\@(\ref{eqn8}). Taking into account that $t_c\propto \kappa$ [Eq.\@(\ref{eqn3}) and Eq.\@(\ref{eq:set4C})], we can divide $t^A_c$ and $t^B_c$ to eliminate the common $t_{xx/yy}$ term and express an unknown chirality parameter, here $\kappa_A$, in terms of a reference one, here $\kappa_B$, as:
\begin{equation}
\label{eqn9}
\kappa_A=\kappa_B\cdot\frac{\thetaTETM^A+i\etaTETM^A}{\thetaTETM^B+i\etaTETM^B}.
\end{equation}
This procedure is illustrated in Fig.\,\ref{fig:fig05} for TE-illumination and two inclusions A and B with chirality parameters $\kappa_A$ and $\kappa_B$, respectively, of the form \,\cite{Condon, Zhao2009, Droulias2020PRB}:
\begin{equation}
\label{eqn10}
\kappa_{A/B}(\omega)=\frac{\omega_{\kappa,A/B}\omega}{\omega^2_{0\kappa,A/B}-\omega^2+i\gamma_{\kappa,A/B}\omega}.
\end{equation}
We emphasize here that, contrary to our previous examples where $\kappa$ is constant, here we consider the (realistic) case of dispersive chirality. In this example, for inclusion A we use $\omega_{\kappa,A} = 2\pi\times 1\times 10^{-3}$\,THz, $\omega_{0\kappa,A} = 2\pi\times 230$\,THz, $\gamma_{\kappa,A} = 2\pi\times 1$\,THz, and, for inclusion B, $\omega_{\kappa,B} = 2\pi\times 1\times 10^{-3}$\,THz, $\omega_{0\kappa,B} = 2\pi\times 225$\,THz, $\gamma_{\kappa,B} = 2\pi\times 2$\,THz, to demonstrate that we can accurately determine the chirality of the (unknown) chiral inclusion A, i.e., the true value of $\kappa_A$, using a chiral inclusion of known chirality parameter ($\kappa_B$) and comparing their respective chiroptical signals, obtained using transmission measurements under TE/TM-illumination. In Fig.\,\ref{fig:fig05} (top row) we present the chiroptical signals $\theta{_{_{\rm{TE}}}}$, $\eta{_{_{\rm{TE}}}}$ for the two, known and unknown, chiral inclusions (here, the metal width of the metasurface is $w = 140$\,nm). We also present in Fig.\,\ref{fig:fig05} (bottom row) the chirality parameters of the two inclusions together with the (accurately) retrieved chirality for the inclusion A using the chiroptical signals, as dictated by Eq.\@(\ref{eqn9}). The same result can be also recovered with TM-illumination (see SM). We emphasize here that, with our example, we demonstrate the validity of Eq.\@(\ref{eqn9}) in providing the entire spectral response of an unknown chiral inclusion even for cases of unknown chiral inclusions with high chirality values (in our example, in the order of $\sim 10^{-3}$ on resonance).\\
\indent The result of Eq.\@(\ref{eqn9}) is particularly practical for experiments and applications as it allows for direct determination of the chirality parameter $\kappa$ of an unknown chiral substance with measurements solely in transmission (and with the aid of a calibration measurement). This eliminates the need for (i) knowing both the reflection and transmission amplitudes and (ii) extracting the effective conductivities in an intermediate step. Most importantly, with this procedure we manage to effectively deconvolve the dispersion of the metasurface from the dispersion of the chiral inclusion and obtain the pure spectral response $\kappa$ of the chiral inclusion.
\section{Complete measurement with elliptically polarized beams}
\noindent The anisotropy of our system allows us to gain access to stronger chiroptical signals by using elliptically polarized incident waves, an additional functionality not feasible in isotropic systems (Ref.\,\cite{Droulias2020NL}). To demonstrate this, we start by parametrizing the elliptically polarized incident wave as:
\begin{equation}
\label{eqn11}
\mathbf{E}_{inc}=
\left(\begin{array}{c}E_{0x}\hat{x} \\ E_{0y}\hat{y}\end{array}\right)
=\left(\begin{array}{c}E_0\cos\phi_{\rm{rot}}\hat{x}\\ E_0\sin\phi_{\rm{rot}}e^{i\phi_{\rm{lag}}}\hat{y}\end{array}\right),
\end{equation}
where the angle $\phi_{rot}$ is the angle between the incident wave's $E$-field and the $x$-axis, and $\phi_{lag}$ tunes the phase-lag between the $x$-, $y$-wave components; $E_0$ is a complex constant. By tuning $\phi_{rot}$ and $\phi_{lag}$ we gain access to any desired polarization. Of course, because the metasurface is anisotropic (birefringent), any incident polarization that is not parallel or vertical to the metal wires (equivalent to the fast/slow axis of typical anisotropic systems) will result in $\theta$, $\eta$ signals, even for $\kappa = 0$; therefore, for $\kappa\neq 0$, the $\theta$, $\eta$ signals may contain achiral background contributions from the metasurface itself that may hinder accurate chirality measurements. In this case, subtraction of the (achiral) background contribution is required and this could be realized using two different approaches: performing measurements with and without the chiral layer (similarly to the work of Ref.\,\cite{Zhao2017}), or implementing a signal reversal that allows us to isolate the signal from the chiral inclusions. Next, we examine both approaches individually.

\subsection{Measurements in transmission with background subtraction}
\vspace{-3.5mm}
\begin{figure}[t!]
\centering
\includegraphics[width=0.98\linewidth]{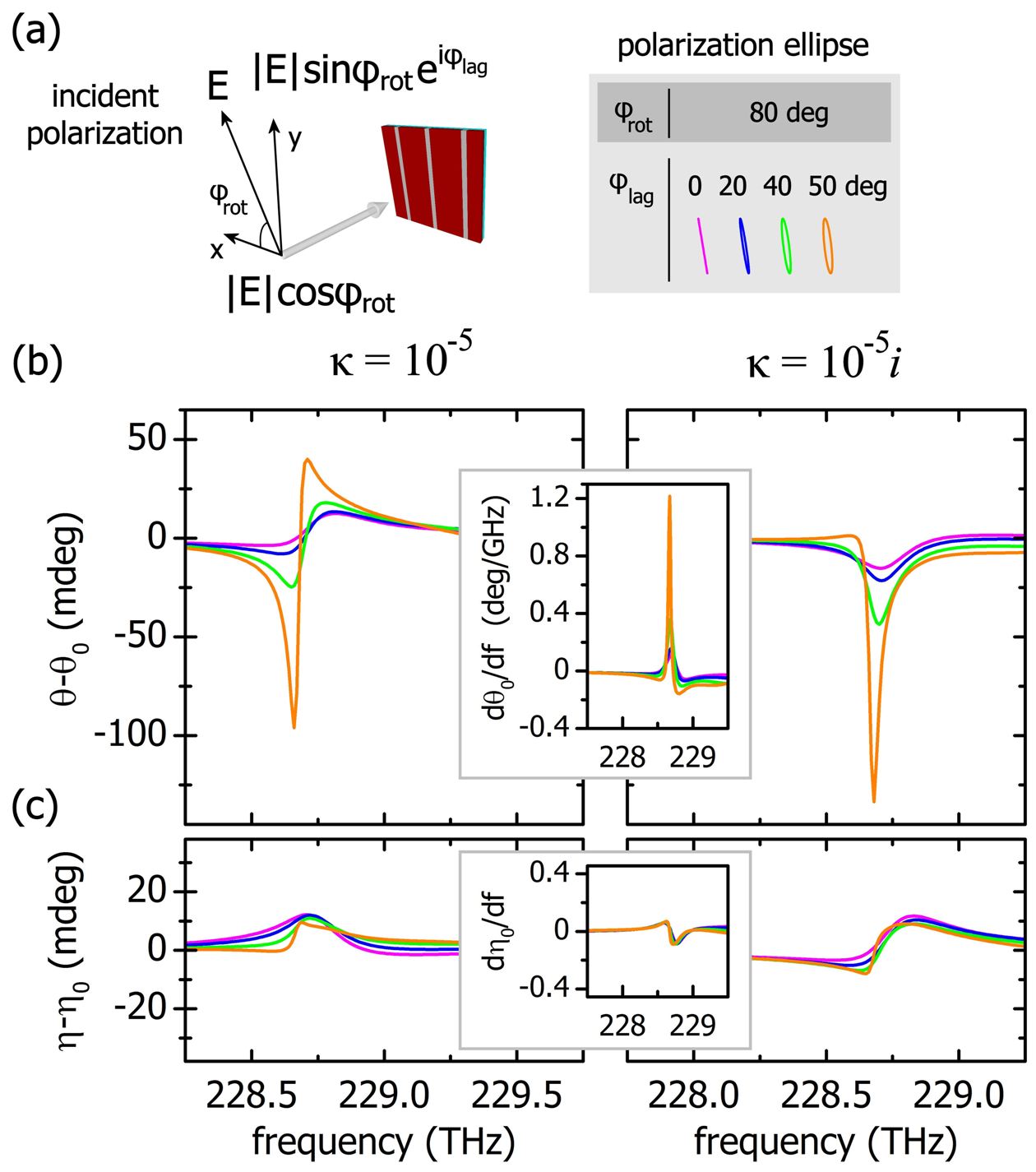}
	\caption{Enhanced chiral sensing using elliptically polarized incident wave. (a) Incident polarization and polarization ellipse for selected examples. Subtraction of far-field measurements of optical rotation and ellipticity in the presence and absence of the chiral layer, i.e., (b) $\theta-\theta_0$, and (c) $\eta-\eta_0$, for the selected polarizations shown in (a); $\kappa = 10^{-5}$ (left panels), $\kappa = 10^{-5}i$ (right panels). For $\phi_{rot} = 80$\,deg, $\phi_{lag} = 50$\,deg, the optical rotation signal is enhanced by a factor of more than $\sim650$ compared to the respective signal of a pure chiral film with $\kappa = 10^{-5}$.}	
    	\label{fig:fig06}
\end{figure}
\noindent In Fig.\,\ref{fig:fig06} we show the first possibility for the system of Fig.\,\ref{fig:fig02} with $w=140$\,nm and a constant chirality parameter $\kappa = 10^{-5}$. In particular, for fixed $\phi_{rot}=80$\,deg, as we increase $\phi_{lag}$ [Fig.\,\ref{fig:fig06}\,(a)] we find that the difference between two measurements with and without the chiral layer, i.e., $\theta-\theta_0$ [where $\theta (\theta_0)$ is with (without) the chiral layer], increases and becomes maximum for $\phi_{lag}=50$\,deg. For this illumination we obtain enhanced chiroptical signals by a factor of more than $\sim650$ compared to the respective signal from the pure chiral film ($\sim 0.14$\,mdeg) [Fig.\,\ref{fig:fig04}\,(c)], as we show in Fig.\,\ref{fig:fig06}\,(b). Additionally, in Fig.\,\ref{fig:fig06}\,(b),\,(c) we demonstrate the ability with such a subtraction procedure to acquire pure chiroptical signals for the case of (purely imaginary) $\kappa = 10^{-5}i$.\\
\indent The strong enhancement of the chiroptical signals [Fig.\,\ref{fig:fig06}\,(b),\,(c)] originates from the anisotropy of the metasurface; under illumination with elliptically polarized beam, the chiroptical signals change as $\theta-\theta_0\sim d\theta_0/d{\rm{f}}$ and  $\eta-\eta_0\sim d\eta_0/d{\rm{f}}$, where ${\rm{f}}$ is the frequency (Ref.\,\cite{Droulias2020NL}). Essentially, the chiroptical signals due to the chiral inclusions are significantly enhanced at frequency ranges where the metasurface’s anisotropy changes abruptly (i.e., the derivatives of $\theta_0,\eta_0$ become large). This is illustrated in the insets of Fig.\,\ref{fig:fig06}, where the calculated derivatives $d\theta_0/d{\rm{f}}$ and $d\eta_0/d{\rm{f}}$ with the chiral layer removed, are proportional to $\theta-\theta_0$ and $\eta-\eta_0$, respectively.

\subsection{Measurements in transmission using a signal reversal: Absolute measurements} \vspace{-3.5mm}
\noindent Removing the chiral inclusion to subtract the background signals may result in unintentional changes in the measurement setup. This can possibly affect the measurement accuracy, and, therefore, it would be useful to be able to perform chirality measurements without the need for sample removal.\\
\indent Our system offers the possibility for absolute measurements by means of a crucial signal reversal with which one can directly isolate the (enhanced) chiroptical signal without the need for sample removal and interference with the system, a unique approach in metamaterial-based chiral sensing schemes: excitation with reversed polarization yields separate polarization effects of opposite sign, thus, enabling the isolation of signals originating only from the chiral inclusion. According to our parametrization of Eq.\@(\ref{eqn11}), this can be realized by reversing $\phi_{rot}$ upon illumination, yielding different chiroptical signals with opposite rotations and ellipticities that we label as $\theta_{\pm}$ and $\eta_{\pm}$. By taking their mean value, i.e., $2\Delta\theta_{rev}\equiv \theta_+ + \theta_-$ and $2\Delta\eta_{rev}\equiv \eta_+ + \eta_-$, any signal originating from the metasurface is cancelled, as well as any other potential achiral backgrounds, while the pure chiroptical signal, which is even under this polarization reversal, doubles. Thus, the importance of the signal reversal becomes apparent: under realistic experimental conditions, one can appropriately tune the frequency of the probing radiation around the resonance of the metasurface and apply this reversal (e.g., with the use of polarization modulators), isolating the chiral signal even under the presence of high-noise environments and other achiral effects (generally, signal reversals have been crucial for enabling sensitive measurements of circular birefringence in conditions where traditional polarimetry fails to perform\,\cite{Sofikitis2014,Bougas2015,Visschers2020}).\\
\indent The average signals $\Delta\theta_{rev}$ and $\Delta\eta_{rev}$ can be expressed as (see SM for derivation):
\begin{widetext}
\begin{subequations}
\label{eq:eqn12}
\begin{align}
        \Delta\theta_{rev}=&{\rm{Re}}(-t_c\frac{c^2_{rot}t_{xx}+s^2_{rot}t_{yy}e^{2i\phi_{\rm{lag}}}}{c^2_{rot}t^2_{xx}+s^2_{rot}t^2_{yy}e^{2i\phi_{\rm{lag}}}}),
        \label{eq:set12A} \\
        \Delta\eta_{rev}=&{\rm{Im}} \Big(-t_c (c^2_{rot}t_{xx}^{\ast}+s^2_{rot}t_{yy}^{\ast})  \frac{|c^2_{rot}t_{xx}^2+s^2_{rot}t_{yy}^2 e^{2i\phi_{\rm{lag}}}|}{(c^2_{rot}|t_{xx}|^2+s^2_{rot}|t_{yy}|^2)^2}
       \nonumber \\
       &+ \frac{{\rm{Re}}\big\{t_c (t_{xx}-t_{yy}) (c^2_{rot}e^{-i\phi_{\rm{lag}}} t_{xx}^2+s^2_{rot}e^{i\phi_{\rm{lag}}} t_{yy}^2)^{\ast}\big\} 2c^2_{rot}s^2_{rot}e^{-i\phi_{\rm{lag}}}t_{xx}t_{yy}^{\ast}} {(c^2_{rot}|t_{xx}|^2+s^2_{rot}|t_{yy}|^2)^2 |c^2_{rot}t_{xx}^2+s^2_{rot} e^{2i\phi_{\rm{lag}}}t_{yy}^2|} \Big) \label{eq:set12B},
\end{align}
\end{subequations}
\end{widetext}
where $c_{rot} = \cos(\phi_{rot})$, $s_{rot} = \sin(\phi_{rot})$ and $t_{xx}$, $t_{yy}$, $t_c$ are the transmission amplitudes as given by Eqs.\@(\ref{eq:set4A})-(\ref{eq:set4C}), i.e., those obtained with TE/TM linearly polarized illumination. Because $t_{xx}$, $t_{yy}$, $t_c$ do not depend on $\phi_{rot}$, $\phi_{lag}$, the form of Eq.\@(\ref{eq:set12A}) implies that there is always a combination of $\phi_{rot}$, $\phi_{lag}$ that can minimize and possibly eliminate the denominator entirely, thus enhancing $\Delta\theta_{rev}$, as we demonstrate in Fig.\,\ref{fig:fig06}. On the other hand, because the denominator in Eq.\@(\ref{eq:set12B}) involves sums of positive quantities, a similar conclusion cannot be drawn directly for $\Delta\eta_{rev}$.
\begin{figure}[t!]
\centering
\includegraphics[width=0.99\linewidth]{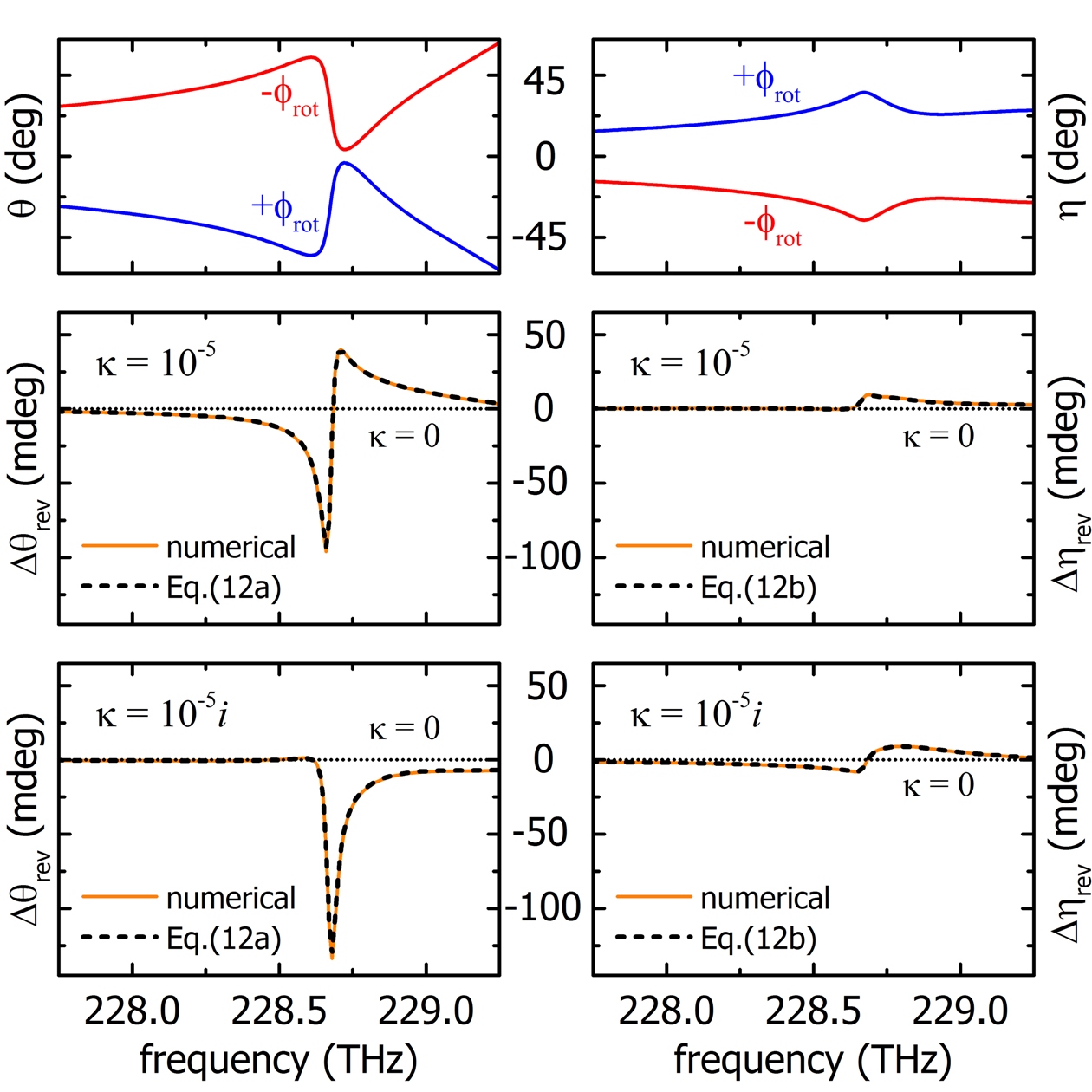}
	\caption{Absolute measurement of real- and imaginary-valued chirality parameter, $\kappa$, with polarization reversal of elliptically polarized incident wave. Measurements for $\phi_{lag} = 50$\,deg with $\phi_{rot} = \pm80$\,deg (blue and red lines, respectively, top row) are averaged to yield pure chiral signals $\Delta \theta_{rev}$, $\Delta \eta_{rev}$ (orange lines, middle and bottom row), for the case of $\kappa = 10^{-5}$ (solid line; middle panels), and $\kappa = 10^{-5}i$ (solid line; lower panels). The analytical plots of Eqs.\@(\ref{eq:set12A}) and \@(\ref{eq:set12B}) for the respective cases are also shown (black dashed lines). Dotted lines: $\kappa = 0$.}	
    	\label{fig:fig07}
\end{figure}
\indent To illustrate these observations, in Fig.\,\ref{fig:fig07} we present simulations of $\theta$ and $\eta$ for incident elliptical waves with $\phi_{rot} = \pm 80$\,deg and $\phi_{lag} = 50$\,deg, which are averaged ($\Delta\theta_{rev}$, $\Delta\eta_{rev}$) to yield signals originating only from the chiral inclusions (we perform the calculations for both cases of purely real- and imaginary-valued chirality parameters, $\kappa = 10^{-5}$ and $\kappa = 10^{-5}i$, respectively). In Fig.\,\ref{fig:fig07} we also present the analytical plots of Eqs.\@(\ref{eq:set12A}) and \@(\ref{eq:set12B}) (dashed lines), for which we use $t_{xx}$, $t_{yy}$ with $\kappa = 0$ and $t_c$ with $\kappa = 10^{-5}$, from the simulations. Note that our results are also in agreement with the measurements we present in Fig.\,\ref{fig:fig06}, where the signals originating from the chiral inclusions are obtained via a background subtraction procedure.

\subsection{Sensitivity on polarization reversal errors}
\vspace{-3.5mm}
\begin{figure}[ht!]
\centering
\includegraphics[width=\linewidth]{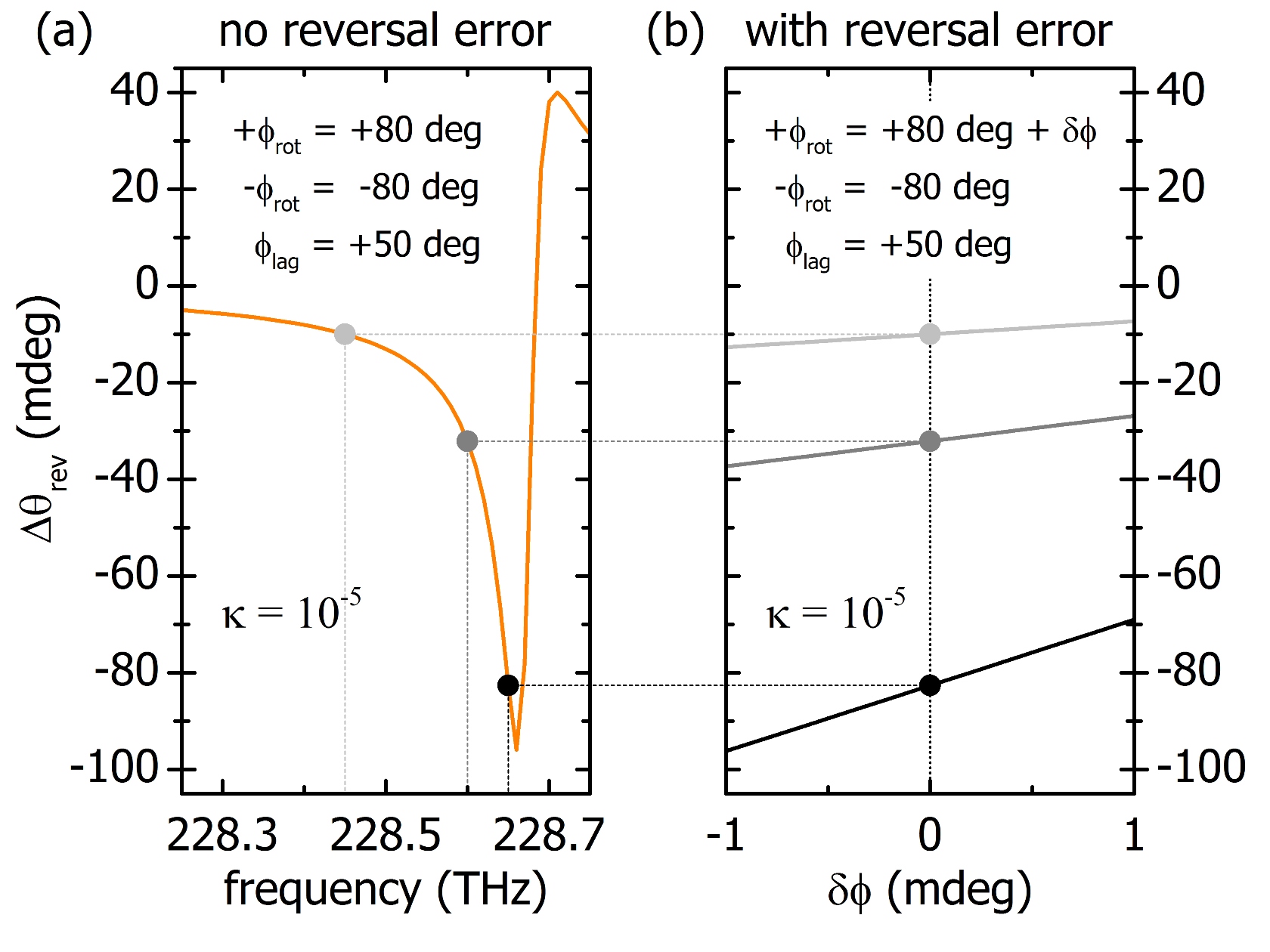}
	\caption{Dependence of chiroptical rotation signal to polarization reversal errors (i.e., measurement sensitivity). (a) Averaged signal $\Delta \theta_{rev}$ for the case of $\kappa = 10^{-5}$ with $\delta \phi = 0$ (b) Averaged signal $\Delta \theta_{rev}$ as a function of $\delta \phi$ for the selected operation points marked in (a).}	
    	\label{fig:fig08}
\end{figure}
\noindent Absolute chiral sensing using the proposed signal reversal, requires the ability to precisely control the polarization state of the incident beam [accurate tuning of the principal axis of the polarization state ($\phi_{rot}$) and the ellipticity ($\phi_{lag}$)] for sensitive measurements. In our scheme, to apply the polarization reversal it suffices to fix $\phi_{lag}$ and then reverse $\phi_{rot}$ as we demonstrate in Figs.\,\ref{fig:fig06}\,\&\,\ref{fig:fig07}, i.e., it is not required to manipulate both $\phi_{rot}$ and $\phi_{lag}$ simultaneously. In practical implementations, however, reversal imperfections may affect the measurement precision.\\
\indent To quantify such a possibility, we introduce a reversal error, $\delta \phi$, as $+\phi_{rot}+\delta \phi$ and $-\phi_{rot}$, and we repeat the measurements of $\Delta\theta_{rev}$ of Fig.\,\ref{fig:fig07} [$\kappa = 10^{-5}$ and $\phi_{rot} = \pm 80$\,deg, $\phi_{lag} = 50$\,deg]. In Fig.\,\ref{fig:fig08}\,(a) we show the chiroptical rotation signal after the application of the signal reversal, $\Delta\theta_{rev}$, for $\delta \phi = 0$, and in Fig.\,\ref{fig:fig08}\,(b) we show $\Delta\theta_{rev}$ as a function of $\delta \phi$ for a few selected operation (frequency) points [as marked in Fig.\,\ref{fig:fig08}\,(a)]. We observe that the measurement's sensitivity in the reversal error $\delta \phi$ increases with signal strength, which is expected as the latter changes as $\sim d\theta_0/d{\rm{f}}$. In particular, for $|\delta \phi|\sim 1$\,mdeg we observe deviations of several mdeg in the measured signal (depending on the operational/measurement frequency), and these increase as we approach the frequency where the metasurface's anisotropy changes abruptly. Notwithstanding, the sensitivity limit of any polarimetric (optical) measurement is ultimately defined by the photon shot noise [which for $\sim$1\,mW of laser radiation at $\sim$1310\,nm ($\sim$229\,THz) is at the sub-$\mu$deg/$\sqrt{\rm{Hz}}$ level] and, even in the case of commercially available optical spectro-polarimeters, polarimetric sensitivities can reach $<100\,\mu$deg levels. As such, signal-reversal control is feasible with high accuracy and precision (i.e., $\delta\phi<0.1$\,mdeg), ensuring the ability to perform sensitive absolute chiral sensing using our proposed signal reversal.  

\subsection{Determination of unknown chirality with measurements in transmission}
\vspace{-3.5mm}
\noindent In experiments it would be particularly useful to have a calibration scheme for measurements with elliptical polarization, similar to the case of TM/TE illumination, where an unknown chirality $\kappa_A$ is determined in terms of a reference chirality $\kappa_B$. Previously, the simple form of Eq.\@(\ref{eqn9}) was derived on the basis that the rotation ($\theta$) and ellipticity ($\eta$) are the real and imaginary part of the same complex quantity, which is proportional to $\kappa$ [see Eqs.\@(\ref{eq:set7A}),\@(\ref{eq:set7B})]. Although under elliptical illumination, $\Delta\theta_{rev}$ and $\Delta\eta_{rev}$ are the real and imaginary part of different complex quantities [see Eqs.\@(\ref{eq:set12A}),\@(\ref{eq:set12B})], as we show in the SM, it is still possible to obtain the unknown chirality in a similar form, as:

\begin{figure}[t!]
\centering
\includegraphics[width=\linewidth]{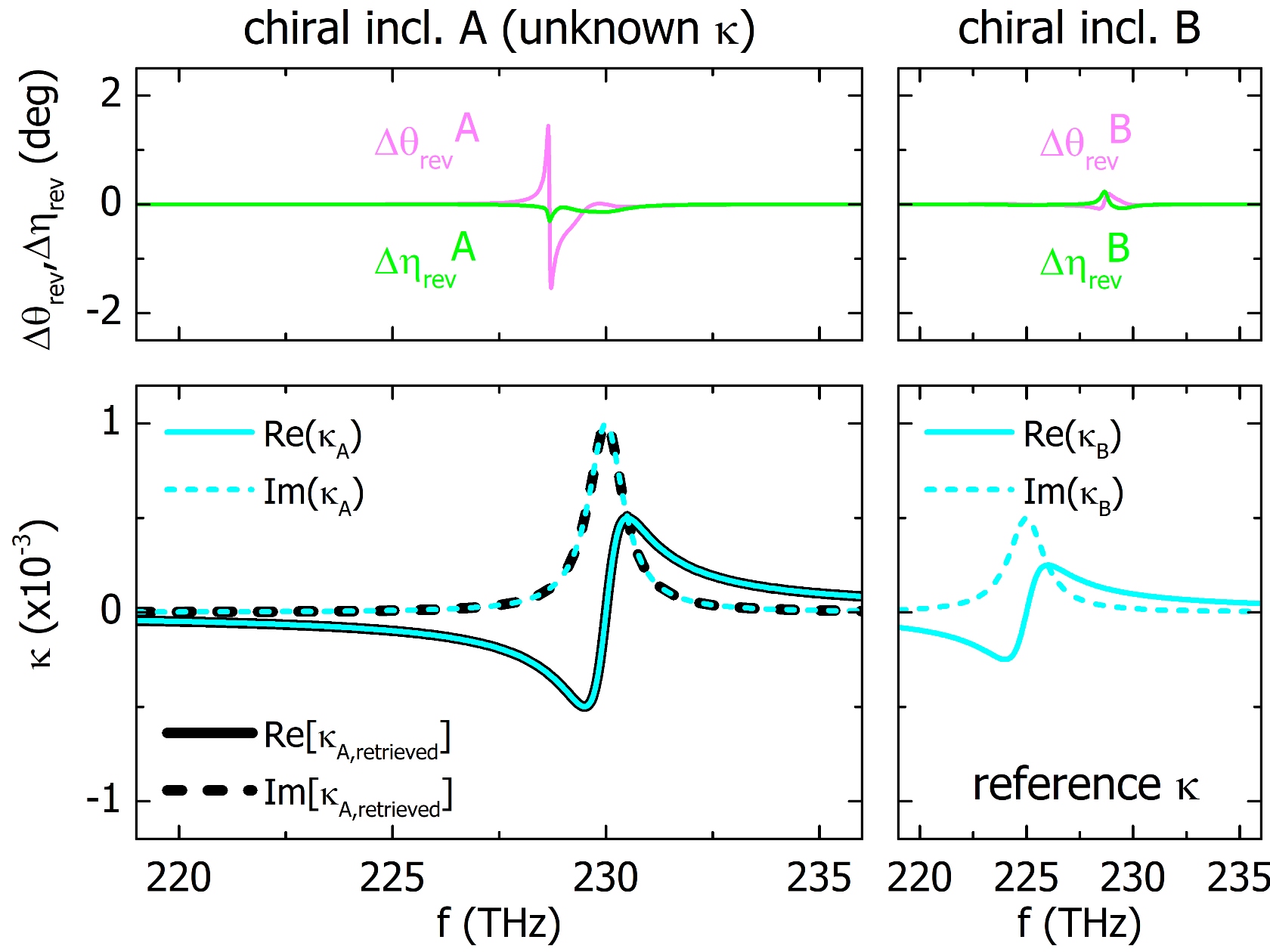}
	\caption{Retrieval of the chirality parameter $\kappa_A$ of an unknown inclusion A, using a reference chirality parameter $\kappa_B$ of inclusion B, and an elliptically polarized beam with $\phi_{rot}=\pm 80$ deg, $\phi_{lag}=50$ deg. $\textit{Top row}$: chirality parameters $\kappa_A$, $\kappa_B$ of the two chiral inclusions (cyan lines) and retrieved $\kappa_A$ (black lines) of inclusion A, using the reference values of $\kappa_B$ and the chiroptical signals $\Delta \theta_{rev}$, $\Delta \eta_{rev}$. $\textit{Bottom row}$: numerically calculated chiroptical signals $\Delta \theta_{rev}$, $\Delta \eta_{rev}$ for each inclusion, for the achiral anisotropic metasurface of Fig. 2 with $w = 140 nm$.}	
    	\label{fig:fig09}
\end{figure}

\begin{equation}
\label{eqn13}
\kappa_A=\kappa_B\frac{\Delta\theta^A_{rev}+g\Delta\eta^A_{rev}}{\Delta\theta^B_{rev}+g\Delta\eta^B_{rev}},
\end{equation}
\noindent where $g$ is a function of the angles $\phi_{rot},\phi_{lag}$ characterizing the elliptical polarization, and the co-transmission amplitudes $t_{xx}, t_{yy}$ obtained previously with TE/TM illumination. 
The result of Eq.\@(\ref{eqn13}) generalizes Eq.\@(\ref{eqn9}) for any arbitrary incident polarization. Indeed, for $\phi_{rot}=0$ (TM illumination) or $\phi_{rot}=\pi/2$ (TE illumination), $g=i$ and Eq.\@(\ref{eqn13}) reduces to Eq.\@(\ref{eqn9}). As an example, in Fig.\,\ref{fig:fig09} we use Eq.\@(\ref{eqn13}) to retrieve the unknown chirality $\kappa_A$ in terms of the reference chirality $\kappa_B$ using elliptically polarized illumination with $\phi_{rot}=\pm 80$ deg, $\phi_{lag}=50$ deg. For $\kappa_A$, $\kappa_B$ we use the parameters previously used in Fig.\,\ref{fig:fig05} to demonstrate the equivalent retrieval with TE illumination. \\
\indent Finally, we wish to add that, with the introduction of an additional reference chirality $\kappa_C$, we can also retrieve the unknown chirality entirely in terms of the $\Delta\theta, \Delta\eta$ signals of the three molecules, i.e., without the need to involve the angles $\phi_{rot},\phi_{lag}$ and the co-transmission amplitudes $t_{xx}, t_{yy}$ (see SM for details).

\begin{figure*}[t!]
\centering
		\includegraphics[width=0.95\linewidth]{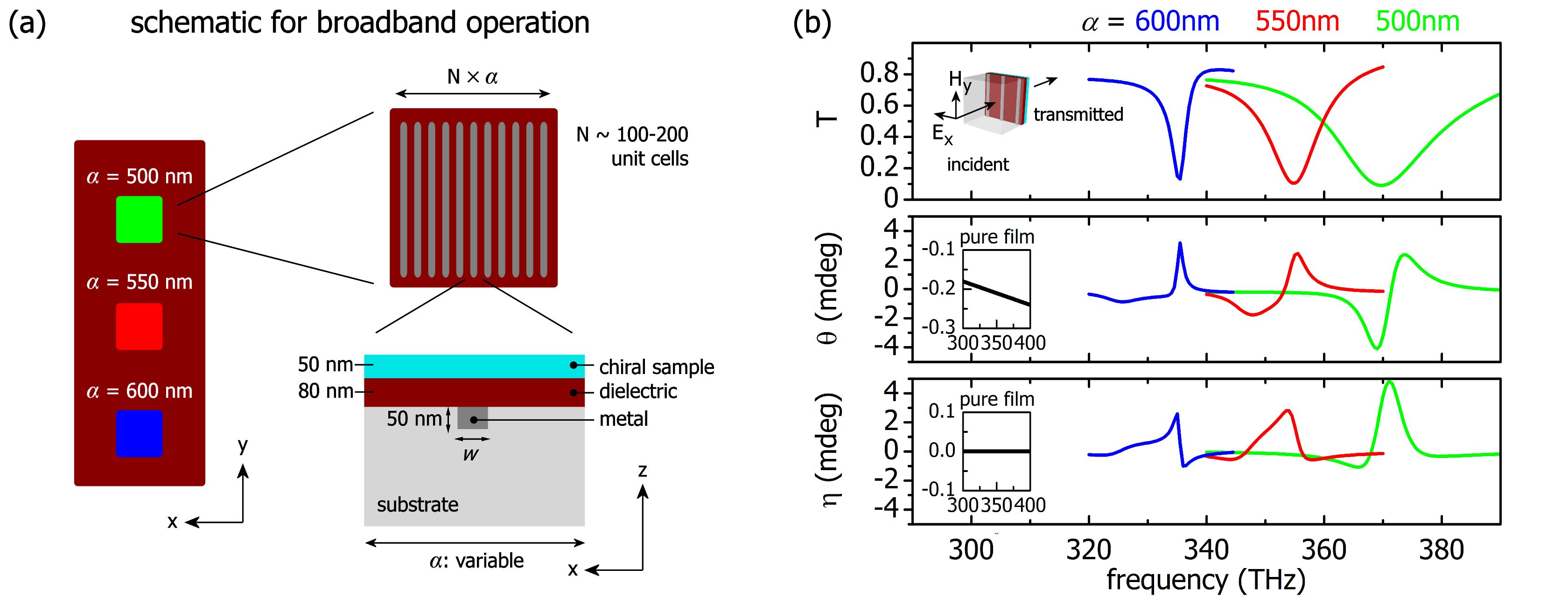}
	\caption{Implementation for broadband operation. (a) Pixelated metasurface composed of a two-dimensional array of meta-units with resonant frequencies tuned over a wide spectral range (here, $\sim$70\,THz) to target, e.g., specific molecular spectral features. (b) Performance for TM-illumination. For this example the metallic wires have constant width $w = 80$ nm and variable periodicity $\alpha$, ranging from 500\,nm to 600\,nm, as shown.}	
    	\label{fig:fig10}
\end{figure*}

\section{Broadband operation}
\noindent The operational spectral range of our metasurface depends on the properties of the utilized resonant modes; altering those, one can selectively tune the metasurface's operation frequency to approach that, for instance, of a target substance. For our proposed design, we can tune the operation of the metasurface over a wide spectral range - from near-ultraviolet (NUV) to near-infrared (NIR) frequencies - coarsely and finely by changing, respectively, the properties of the materials forming the metasurface and its geometric characteristics. Particularly, in Ref.\,\cite{Droulias2020NL} we have chosen specific metasurface design parameters for operation at visible frequencies [370\,THz (800\,nm)], while in this work we select design parameters for operation at near-infrared frequencies [230\,THz (1300\,nm)].\\
\indent Considering this general design principle we can see how our system can be implemented for chiroptical spectroscopy over different spectral ranges with the capacity for wide spectral coverage. For this reason, we present here two particular designs: (i) one suitable for broadband spectral coverage in the visible and near-infrared on a single platform, and (ii) one suitable for operation at near-ultraviolet frequencies.

\begin{figure*}[ht!]
\centering
		\includegraphics[width=0.95\linewidth]{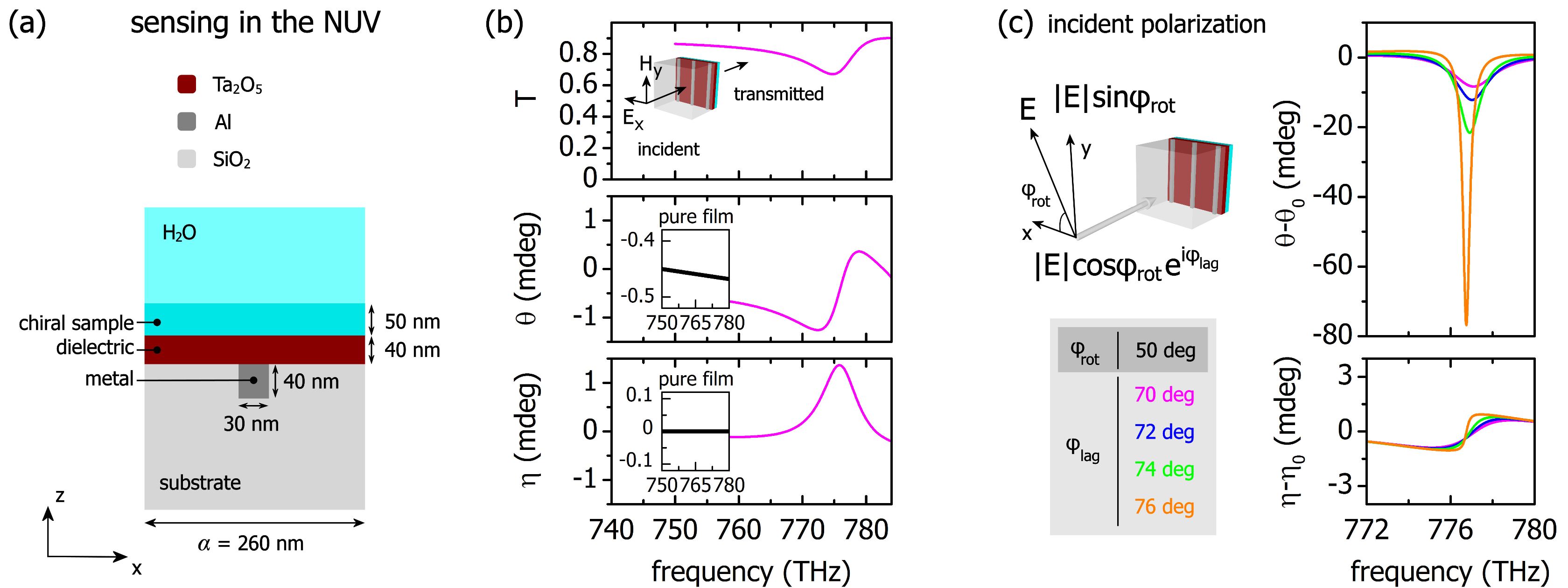}
	\caption{Design for operation in the near-ultraviolet (NUV). (a) Sample schematic. (b) Performance for TM-illumination. (c) Performance for illumination with elliptical beam. For TM illumination the chiroptical signals are enhanced by a factor of 2-3 [compared to single-pass transmission polarimetric measurements; insets (b)], while under elliptical-illumination the chiroptical rotation signal ($\theta$) is enhanced by a factor of $\sim$165.}	
    	\label{fig:figB}
\end{figure*}
\subsection{Pixelated sensor for operation in the visible \& near-infrared}
\vspace{-3.5mm}
\noindent For broadband spectral coverage in the visible and near-infrared we visualize a pixelated chiral sensor, as shown in Fig.\,\ref{fig:fig10}\,(a), where each \textit{meta-pixel} is dedicated to the detection of chirality at a particular spectral range, following a similar principle of operation as the recently demonstrated imaging-based molecular barcoding enabled by high-Q dielectric resonators for the detection of mid-infrared molecular fingerprints\,\cite{Tittl2018}. To develop such a sensor we start by selecting a design for operation at a particular frequency range, and to span a broad spectral range, we finely tune the design parameters to change the resonant frequencies of the TE$_{20}$ and TM$_{20}$ modes assigning, thus, a new meta-pixel to the overall design. By repeating this procedure we can cover practically any desired spectral range.\\
\indent As an example, in Fig.\,\ref{fig:fig10} we demonstrate a system for operation at $\sim 770-920$ nm (spanning a range of $\sim 70$\,THz). In particular, we start with designing the system for operation at 800\,nm (374\,THz) and we use a slab with refractive index $n_{\rm{slab}}\!=\!2.52$ (e.g. TiO$_2$) and thickness $t_d =80$\,nm on a glass substrate with $n = 1.45$. The metallic wires are made of silver (Ag) and are placed on one side of the dielectric slab, as illustrated in the schematic of Fig.\,\ref{fig:fig10}\,(a)\,\cite{Droulias2020NL}. The wire cross-section on the $xz$-plane has thickness 50\,nm and width 80\,nm, and the wire periodicity is $\alpha = 500$\,nm. For the simulation we choose a chiral layer, which we place on the slab, of thickness $t_c = 50$\,nm and refractive index $n_c\!=\! 1.33-10^{-4}i$, and the entire space above the chiral layer is water with $n = 1.33$. For TM excitation we obtain chiroptical rotation ($\theta$) signals as large as 6.5\,mdeg peak-to-peak, for a transmittance of $\sim$10\%. As a comparison, the optical rotation signal from a transmission measurement of a 50\,nm chiral layer with $\kappa=10^{-5}$ at 800\,nm is $\sim$0.24\,mdeg, achieving, thus, enhancements by a factor of $\sim$27 [Fig.\,\ref{fig:fig10}\,(b), insets]. For TE illumination or excitation with elliptical beam we can further enhance our signals, even by an order of magnitude, as we show in Ref.\,\cite{Droulias2020NL}. \\
\indent To shift the operation at lower frequencies, we increase the periodicity of the metallic wires; by increasing the periodicity we shift the resonant frequencies of the electric and the magnetic mode to lower frequencies, and by changing the cross-section of the metallic wires we can fine-tune the spectral overlap of the two modes. To demonstrate the versatility of our design, here we only change the periodicity $\alpha$ (all other material and geometrical parameters remain unchanged). In particular, for $\alpha = 550$\,nm we can lower the operation by 20\,THz (at $850$\,nm) and for $\alpha = 600$\,nm we can further lower the operation by an additional 20\,THz (at $900$\,nm). We see that, even for the case of solely tuning the wire periodicity, we can cover a broad spectral range and support chiroptical rotation ($\theta$) signals as large as 4\,mdeg peak-to-peak, for a transmittance of $\sim10$\%.\\
\indent Given the design shown in Fig.\,\ref{fig:fig10}\,(a), then, for 100-200 unit cells per meta-pixel, and for a range of wire periodicities of, e.g., $\alpha = 450\,{\rm{nm}} - 700$\,nm, we can develop an overall design that enables chiroptical molecular spectroscopy over a broad spectral range (of more than 100\,THz) on a sub-mm$^2$ device.

\subsection{Near-ultraviolet (NUV) operation} \vspace{-3.5mm}
\noindent Most recent nanophotonic/metamaterial-based chiral sensing platforms have been largely designed to operate at infrared and visible frequencies where materials (metals/dielectrics) have (typically) low losses and, hence, most experimental demonstrations have been performed on large molecules that possess chiroptical bands at visible and infrared frequencies. However, for most (small) molecules, strong chiral absorption features are in the ultraviolet (UV) spectral region, and the ability to improve the performance of chiral sensing platforms in the UV can be pivotal\,\cite{Hu2020}.\\
\indent We present in Fig.\,\ref{fig:figB} a specific design of our metasurface platform that allows for enhanced chiroptical spectroscopy at near-UV frequencies ($\sim$380\,nm). To achieve operation at the desired spectral region we modify the design of our previous example (Fig.\,\ref{fig:fig10}) by replacing the slab and metallic wire material with Ta$_2$O$_5$ and Al, respectively, and by reducing the slab thickness to 40\,nm and the wire periodicity to 260\,nm; the wire cross-section on the $xz$-plane is modified to have thickness 40\,nm and width 30\,nm. The 50\,nm chiral film is characterized by $\kappa = 10^{-5}$ and $n = 1.339-10^{-4}i$ (for water $n=1.339$ at $\sim$400\,nm) and the glass substrate has $n = 1.47$ [Fig.\,\ref{fig:figB}\,(a)]. Despite the relatively weak enhancement for TM-illumination [Fig.\,\ref{fig:figB}\,(b)], we find that under elliptical illumination [Fig.\,\ref{fig:figB}\,(c)], the chiroptical rotation signal ($\theta$) is enhanced by a factor of $\sim$165, compared to the respective signal for measuring the chiral film without the metasurface.\\
\indent With pursuing further increase in frequency (from the near-UV to the UV), the choice of materials for the implementation of the proposed scheme becomes a challenging task. The reason is that in the UV several dielectrics become lossy and, in particular, many metals lose their metallic character. Other than that, and as we demonstrate with this example of operation at near-UV frequencies, our proposed scheme is general and there is no fundamental limitation with respect to the operational spectral range.

\section{Conclusions}
\vspace{-3.5mm}
\noindent In this work we analyzed the key functionalities and benefits of our proposed scheme for metasurface-based enhanced chiral sensing, which is based on the anisotropic response of the metasurface, rather than the superchirality of the generated near-fields, as in most contemporary nanophotonic-based approaches. Our design provides a platform where aspects crucial for chiral sensing in the nanoscale can be realized: (i) enhanced chiroptical signals by more than two orders of magnitude for ultrathin, subwavelength, chiral samples over a uniform and accessible area, (ii) complete measurements of the total chirality (magnitude and sign of both its real and imaginary part), and (iii) measurements in an absolute manner, i.e., without the need for sample removal, due to the possibility for a crucial signal reversal (excitation with reversed polarization). We derived analytically, and verified numerically, simple formulas that provide insight to the sensing mechanism and explain how our system offers additional degrees of freedom with respect to other isotropic metasurfaces. Importantly, our theoretical analysis provides a unified description for achiral anisotropic metasurfaces for chiral sensing that extends beyond the specific example systems we have used. We demonstrated practical measurement schemes for the complete and unambiguous determination of an unknown chirality and we provided the design principles towards broadband operation - from near-infrared to near-ultraviolet frequencies - opening the way for highly sensitive nanoscale chiroptical spectroscopy.

\section*{Acknowledgement}
\vspace{-3.5mm}
\noindent We acknowledge the support of the European Commission Horizon 2020 ULTRACHIRAL Project (grant no. FETOPEN-737071).

\bibliography{main}
\end{document}